\documentclass[12pt,preprint]{aastex}

\shorttitle{}
\shortauthors{Usui et al.}

\begin{document}

\title{ Albedo Properties of Main Belt Asteroids Based on the Infrared All-Sky Survey of the Astronomical Satellite {\it AKARI} }
\author{
Fumihiko  Usui\altaffilmark{1}, 
Toshihiro Kasuga\altaffilmark{2}, 
Sunao     Hasegawa\altaffilmark{1}, 
Masateru  Ishiguro\altaffilmark{3}, 
Daisuke   Kuroda\altaffilmark{4}, 
Thomas G. M\"{u}ller\altaffilmark{5}, 
Takafumi  Ootsubo\altaffilmark{6},  
and
Hideo     Matsuhara\altaffilmark{1}
}
\email{usui@ir.isas.jaxa.jp}

\altaffiltext{1}{Institute of Space and Astronautical Science, 
Japan Aerospace Exploration Agency, 
3-1-1 Yoshinodai, Chuo-ku, Sagamihara 252-5210, Japan}

\altaffiltext{2}{Public Relations Center, 
National Astronomical Observatory of Japan, 
2-21-1 Osawa, Mitaka, Tokyo 181-8588, Japan}

\altaffiltext{3}{Department of Physics and Astronomy, 
Seoul National University, 
San 56-1, Shillim-dong Gwanak-gu, Seoul 151-742, South Korea}

\altaffiltext{4}{Okayama Astrophysical Observatory, 
National Astronomical Observatory, 
3037-5 Honjo, Kamogata-cho, Asakuchi, Okayama 719-0232, Japan}

\altaffiltext{5}{Max-Planck-Institut f\"{u}r Extraterrestrische Physik, 
Giessenbachstra\ss e, 85748 Garching, Germany}

\altaffiltext{6}{Astronomical Institute, Tohoku University, 
6-3 Aoba, Aramaki, Aoba-ku, Sendai 980-8578, Japan}

\begin{abstract}

We present an analysis of the albedo properties of main belt asteroids detected by 
the All-Sky Survey of the infrared astronomical satellite {\it AKARI}. The characteristics of 5120 asteroids 
detected by the survey, including their sizes and albedos, were cataloged in 
the Asteroid Catalog Using {\it AKARI} (AcuA). 
Size and albedo measurements were based on the Standard Thermal Model, using inputs of infrared fluxes
and absolute magnitudes measured at optical wavelengths.
Main belt asteroids, which account for 4722 of the 5120 AcuA asteroids, 
have semimajor axes of 2.06 to 3.27 AU, except for the near-Earth asteroids. 
AcuA provides a complete data set of all main belt asteroids brighter than the absolute magnitude of $H < 10.3$, 
which corresponds to the diameter of $d > 20$ km. We confirmed that 
the albedo distribution of the main belt asteroids is strongly bimodal as was already known from the past observations, 
and that the bimodal distribution occurs not only in the total population, but also within inner, middle, 
and outer regions of 
the main belt. The bimodal distribution in each group consists of low-albedo components in C-type asteroids 
and high albedo components in S-type asteroids. 
We found that the small asteroids have much more variety in albedo than the large asteroids. 
In spite of the albedo transition process like space weathering, 
the heliocentric distribution of the mean albedo of asteroids in 
each taxonomic type is nearly flat. 
The mean albedo of the total, on the other hand, gradually decreases with an increase in 
semimajor axis. This can be 
explained by the compositional ratio of taxonomic types; that is, the proportion of dark asteroids such as C- and D-types 
increases, while that of bright asteroids such as S-type decreases, with increasing heliocentric distance.
The heliocentric distributions of X-subclasses: E-, M-, and 
P-type, which can be divided based on albedo values, are also examined. P-type, 
which is the major component in X-types, are distributed throughout the main belt regions, 
and the abundance of P-type increases beyond 3 AU. This distribution is similar to that of C- or D-types. 

\end{abstract}


\section{Introduction}

Studies of the physical properties of asteroids are of fundamental importance to our understanding
of the origin, evolution, and structure of the solar system. 
In order to investigate the compositions of asteroids and the chemical and thermal processes of alternations, 
it is essential to examine detailed mineralogical characterizations of individual asteroids 
(e.g.,~\citet{Gaffey2002}). Taxonomic classifications of asteroids 
represent a broader, alternative approach for appraising the compositions and surface conditions 
for large numbers of asteroids.
As is commonly accepted (and first explicitly described by ~\citet{Chapman1975}), most asteroids 
can be classified as either C- or S-type. C-type asteroids are historically associated 
with carbonaceous chondrites, and S-type asteroids with stony-iron meteorites. 

Several outstanding works have defined methodologies for the taxonomic classification of 
asteroids (e.g., ~\citet{Tholen1984, Bus1999, Lazzaro2004, Carvano2010}). 
These taxonomies are based on statistically processed information 
parameterizing ground-based spectroscopic observations within optical and near-infrared wavelengths. 
One of the notable results of these taxonomic classifications, as presented by ~\citet{Bus2002}
(and also by \citet{Mothe-Diniz2003}) is the non-uniform heliocentric clustering of taxonomic types in 
main belt regions; asteroids of each taxonomic type (C, S, X, D + Ld + T, and V)
show bias-corrected distributions at heliocentric distances of 2.1 to 3.3 AU. 
The interpretation of these results is that S-, C-, and D-type asteroids are distributed 
progressively further from the sun in that order, 
and that the bodies located further from the sun are less affected by metamorphism and contain 
a greater proportion of primitive materials than those closer to the sun
(i.e., S-type asteroids are the most metamorphosed and contain the least amount of primitive 
material and D-type asteroids are the least metamorphosed and contain the greatest proportion 
of volatile material). 

Albedo data, along with taxonomic classifications of asteroids, also 
contribute to our understanding of the large-scale distribution of asteroid compositions. 
Albedo values are strongly dependent on the surface conditions and compositions of asteroids.
The relationship between taxonomic types and albedo is, on the other hand, complex, 
and type determinations cannot be made on the basis of albedo values alone; 
the albedos of C- and S-type asteroids vary widely, even though the albedo of C-types is generally 
low and the albedo of S-types is generally high 
(e.g., ~\citet{Zellner1976}). 

The most direct way to measure the albedo of asteroids is by in-situ observations using flyby and rendezvous missions
(e.g., the spacecrafts {\it Galileo} ~\citep{Johnson1992} for (951) Gaspra in 1991 and (243) Ida/Dactyl in 1993; 
{\it NEAR Shoemaker} ~\citep{Cheng1997} for (253) Mathilde in 1996 and (433) Eros in 2000; 
{\it Deep Space 1} ~\citep{Rayman2003} for (9969) Braille in 1999; 
{\it Stardust-NExT} ~\citep{Brownlee2003} for (5535) Annefrank in 2002; 
{\it New Horizons} ~\citep{Stern2003} for (132524) APL in 2006; 
{\it Rosetta} ~\citep{Glassmeier2007} for (2867) Steins in 2008 and (21) Lutetia in 2010; 
{\it Dawn} ~\citep{Russell2004} for (4) Vesta in 2011--2012). 
Moreover, the sample return from asteroids (by {\it Hayabusa} from (25143) Itokawa in 2005; ~\citet{Fujiwara2006}) 
can provide materials for analytical laboratory determinations of asteroid compositions. 
The two most important physical characteristics of asteroids, albedo and size, are 
coupled by their relationships with the absolute magnitude (e.g., ~\citet{Fowler1992}); 
given the size of an asteroid, the albedo value can be uniquely derived on the basis of 
its absolute magnitude.

Several imaging techniques have been developed to determine the shape of asteroids. 
The most straightforward approach is by direct imaging of the asteroid, for example, by use of the
{\it Hubble Space Telescope} ~\citep{Li2010} or large ground-based telescopes with adaptive optics
~\citep{Drummond2009}. 
Radar observations ~\citep{Ostro2000} and speckle interferometry ~\citep{Drummond1985} are also useful for resolving
the shapes of asteroids, as are stellar occultations
~\citep{Durech2011}; the latter is a powerful approach, which
relies upon well-organized campaign observations by many participants including amateur astronomers. 
Although such methods are readily available, 
they require the convergence of critical conditions, such as the selection of large targets with 
trajectories approaching the earth, and/or narrow observational windows. The sheer number of 
asteroids poses yet another difficulty; as of 2012, the number of known asteroids 
is more than 595,000 (see e.g., 
Minor Planet Center\footnote{http://www.minorplanetcenter.net} at the Smithsonian Astrophysical Observatory), 
a number which precludes detailed observations of all individual bodies.

One of the most effective indirect methods for measuring the sizes and albedos of asteroids is by radiometry, 
which uses the combined flux measurements at thermal infrared and reflected optical wavelengths.
The radiometric technique for determining the sizes and albedos 
of asteroids was developed in the early 1970s using ground-based 
observatories ~\citep{Allen1970, Allen1971, Matson1971}, and the approach has
yielded a wealth of information both on individual objects and large populations of asteroids. 
Using radiometric measurements, a large number of objects can be observed in a short period
of time, thus providing uniform data for large, relatively unbiased populations within the asteroid belt. 
Although radiometry requires careful calibration, once 
calibrated, this method can obtain ``wholesale'' highly accurate measurements of the physical properties of 
large numbers of asteroids. 

Infrared observations using ground-based observatories, especially in the mid-infrared range, 
are severely limited 
on account of moisture conditions in the telluric atmosphere; however, infrared measurements 
obtained using space-borne telescopes are very reliable. 
Furthermore, when integrated into an all-sky survey, large numbers of infrared images can be 
obtained rapidly; moreover, the data are unbiased and uniform. 
The first systematic survey of asteroids using a space 
telescope was made by the {\it Infrared Astronomical Satellite} ({\it IRAS}; ~\citet{Neugebauer1984}), 
which cataloged the sizes and albedos of 2470 asteroids ~\citep{Tedesco2002}. 
Recently, the asteroid catalog was updated with data from 
the infrared astronomical satellite {\it AKARI} All-Sky Survey 
~\citep{Murakami2007}. The Asteroid Catalog Using {\it AKARI} (AcuA) contains an inventory of the sizes and albedos of 
5120 asteroids ~\citep{Usui2011}. 
The {\it Wide-field Infrared Survey Explorer} ({\it WISE}; ~\citet{Wright2010}) also performed an infrared 
all-sky survey and the processed data including large amount of 
asteroid information ({\it WISE}/NEOWISE; ~\citet{Mainzer2011}) are being compiled. 

This paper focuses on the albedos of the main belt asteroids (MBAs) 
measured by {\it AKARI}. 
The spatial distribution of compositions among MBAs is of particular interest, 
because the main belt is the largest reservoir of asteroids in the solar system; 
more than 90\% of asteroids with known orbital elements are classified as MBAs.
Asteroids are thought to be the remnants of planetesimals formed in the early solar system
~\citep{Bottke2002a}. Some were formed near to their current locations, and others 
have migrated from their original birthplaces in conjunction with the migrations of giant planets 
(e.g., \citet{Levison2009}). 
Because of cataclysmic events in the history of the solar system, present-day asteroids, especially MBAs, are 
well mixed and represent multiple origins;
they are confined to certain regions on account of resonance 
effects, and/or have been broken or segmented by mutual collisions. 
Some asteroids originally formed in volatile-poor regions (i.e., the inner parts of the solar system), 
whereas others formed in volatile-rich regions (i.e., the outer parts of the solar system, beyond 
the snow line at the times of formation).
Processes of dynamical evolution and chemical processing may have affected the physical conditions, 
and hence the albedo properties, of asteroid surfaces. 

The main purpose of this study is 
to examine the size dependencies, frequency distributions, and heliocentric distributions 
of the albedos of AcuA MBAs, by comparing to recent taxonomic classifications. 
Famed results about the heliocentric distribution of taxonomic types are presented in \citet{Bus2002}
and \citet{Mothe-Diniz2003}.
They performed a bias correction method based on ~\citet{Zellner1979} and ~\citet{Bus1999}, that is, 
asteroids are divided into several zones according to the semimajor axis and the absolute magnitude, 
and the ratio between the total number of asteroids and the number of classified asteroids in each zone 
are determined, then this ratio is used as the bias correction factor for estimating the fraction of taxonomies 
of unclassified objects under the assumption of size and albedo based on the IRAS data set. 
AcuA, which is based on an all-sky survey 
that lasted more than one year, provides a complete data set of 
all known asteroids brighter than the absolute magnitude 
of $H < 10.3$ for MBAs, which correspond to the size for $d > 20$ km. 
When the objects larger than 20 km are used for analysis, no 
assumptions about size or albedo are needed.
Thus we have carried out no bias correction for the magnitude incompleteness
in our examination of the albedo properties of MBAs.
The paper is organized as follows. In Section 2, we describe
the features of the AcuA data base and the 
division of the main belt into three regions: inner, middle, and outer. 
In Section 3, we describe investigations of albedo properties of MBAs in the context of taxonomic classification. 
In Section 4, we discuss the reasons for the albedo dependencies which are described in Section 3. 
Finally, Section 5 provides a summary.

\section{AcuA main belt asteroids}
\subsection{The AcuA archival data base}
\label{AcuA data}
{\it AKARI} ~\citep{Murakami2007} surveyed more than 96\% of the sky 
during the 18-month course of the cryogenic cooled phase from 2006 to 2007. 
Data were obtained for 6 bands in the mid- to far-infrared 	
spectral range, with a solar elongation of 90$^\circ$ $\pm$ 1$^\circ$.
The mid-infrared part of the All-Sky Survey was conducted at two broad bands centered at 9 \micron ~({\it S9W}) and 
18 \micron ~({\it L18W}), using an on-board InfraRed Camera (IRC; \citet{Onaka2007}). 
The 5$\sigma$ detection limits at {\it S9W} and {\it L18W} were 50 and 90\,mJy, 
respectively, and the spatial resolution of the IRC in the 
All-Sky Survey mode was less than $\sim$10$^{\prime\prime}$ per pixel ~\citep{Ishihara2010}. 
Point sources were extracted from the IRC All-Sky Survey image data, 
numbering 4.8 million at {\it S9W} and 1.2 million at {\it L18W}.
The sources which are detected more than twice are compiled in 
the IRC Point Source Catalog (IRC-PSC;~\citet{Ishihara2010}); however, 
about 20\% of the extracted sources were not contributed for IRC-PSC and were remained unused 
because of a lack of multiple detection at the same position of the sky. 
Asteroids were included in these ``residual'' sources, due to their motions 
against the background sky. By matching the data with the positions of known asteroids, 
about 7000 point sources at {\it S9W} and 14000  point sources at {\it L18W} were identified as asteroids, corresponding to
5120 objects without duplications. 

The radiometric diameters ($d$) and geometric albedos in the visible wavelengths ($p_{\rm v}$) 
of the identified asteroids were calculated using 
the Standard Thermal Model ~\citep{Lebofsky1986} with adjusted beaming parameters of 
$\eta=0.87$ for {\it S9W} and 0.77 for {\it L18W} to reduce the total systematic errors. 
Taking into account several uncertainties, such as the values of the absolute magnitude and the phase slope parameter, 
we computed typical uncertainties in the diameters of asteroids to be about $4.7\%$, 
and those in the geometric albedo results to be about $10.1\%$.
The {\it AKARI} asteroid catalog developed on the basis of these analyses contains data on the sizes and albedos of 5120 asteroids. 
The catalog is named AcuA\footnote{The data are available at http://darts.jaxa.jp/ir/akari/catalogue/AcuA.html}, 
which is the acronym for Asteroid Catalog Using {\it AKARI}. 
More details are described in ~\citet{Usui2011}. 

\begin{figure*}
\begin{center}
\begin{tabular}{c}
(a)\\
\resizebox{95mm}{!}{\epsscale{1.0} \plotone{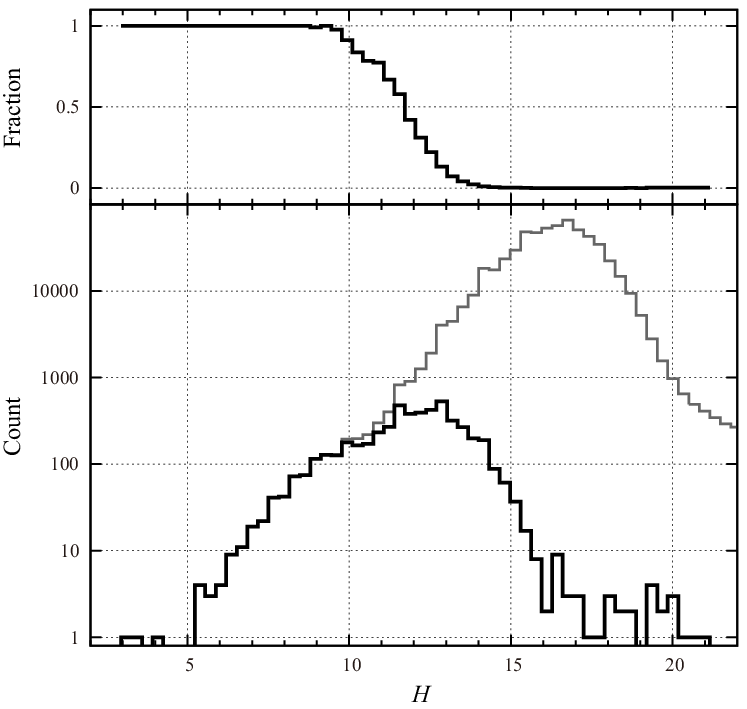}}\\
\\
(b)\\
\resizebox{95mm}{!}{\epsscale{1.0} \plotone{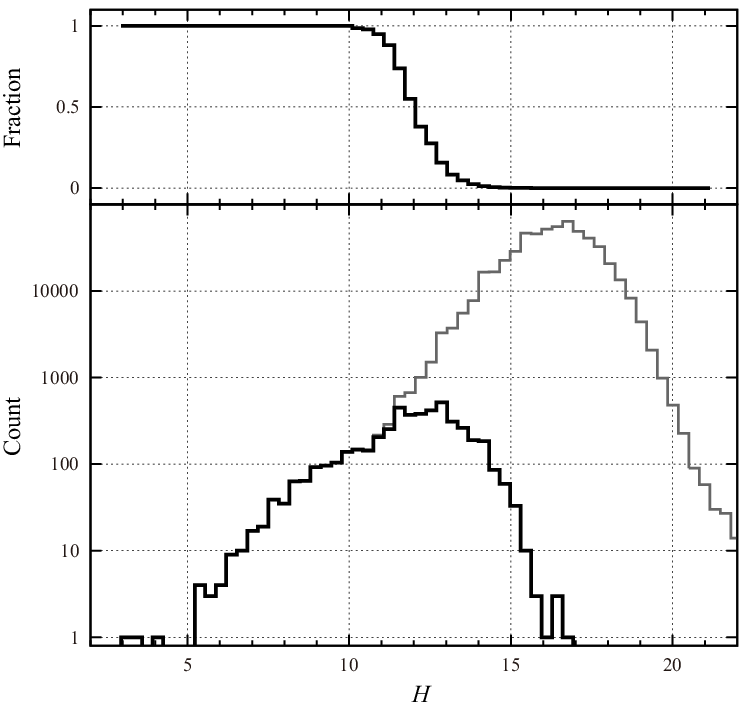}}\\
\end{tabular}
\end{center}
\caption{%
Distribution of absolute magnitude ($H$) for asteroids detected by 
\emph{AKARI} with known orbits. (a) total asteroids with semimajor axes smaller than 6 AU, 
(b) total MBAs. 
The upper panel of each figure shows the fraction of detected asteroids of the total known asteroids, 
and the lower panel shows the distribution of detected asteroids (black line) and that of 
total asteroids with known orbits(gray line). 
The bin size is set to 60 segments for $H$ range of 2 to 22.
} 
\label{Hmag histogram}
\end{figure*}

Figure \ref{Hmag histogram} shows the distribution of $H$ for the AcuA asteroids. 
This figure can be interpreted as reflecting 
the completeness of detections of known asteroids with size and albedo data. 
AcuA, which was constructed based on 16 months of all-sky survey data, 
provides a complete data set of all asteroids brighter than 
absolute magnitude of $H < 9$
within the semimajor axis of $a < 6$ AU, and $H < 10.3$ for all MBAs. 
$H < 10.3$ for MBAs corresponds to $d > 20$ km in size. 
Thus, all AcuA MBAs with values of $d > 20$ km (a total of 1974) are mainly used in the following discussion. 

Based on more recent studies, the uncertainties of the thermal model calculation in ~\citet{Usui2011} 
can be reevaluated. \citet{Pravec2012} calculated the absolute magnitudes 
for 583 asteroids from their dedicated photometric observations over thirty years. 
{When compared with the existing database, the difference in absolute magnitude }
between the database values and their new estimation is about 0.08 within the range of $H < 10.3$. 
The divergence of the slope G parameters are also given in \citet{Pravec2012} as about 0.083, 
though this has little influence on the infrared fluxes, at less than one percent.
The variation of the beaming parameter is given as 0.157 from ~\citet{Masiero2011}. 
These lead to new estimates in the typical uncertainties of AcuA 
in the diameters to 13.6\%, and those in the albedo to 28.1\%.

\subsection{Divisions of the main belt asteroids}
\begin{figure*}
\epsscale{1.0} \plotone{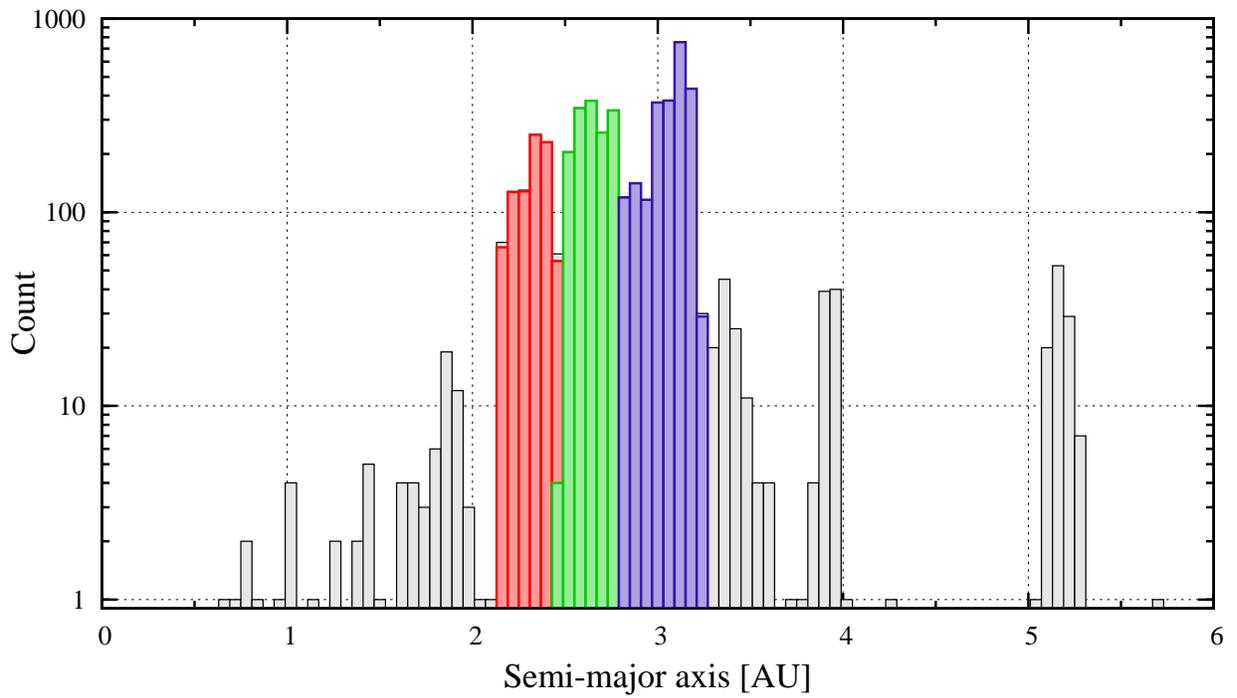}
\caption{Histogram of the number of detected asteroids with {\it AKARI} as a function of the semimajor axis. 
Red, green, and blue boxes denote 
inner, middle, and outer region MBAs, respectively. 
The bin size is set to 100 segments for the semimajor axis range of 0 to 6 AU.
} 
\label{semj count}
\end{figure*}

The number of asteroids inventoried in the AcuA is summarized in Table \ref{number of asteroids}.
Determinations of each taxonomic type are based on 
~\citet{Tholen1984, Bus1999, Lazzaro2004, Carvano2010}, and 
other references summarized in ~\citet{Usui2011}, appendix 4. 
Figure \ref{semj count} shows histograms of the numbers of AcuA asteroids as a function of the semimajor axes. 
We focus on three regions of the main belt which are
defined on the basis of semimajor axis: the inner ($2.06 < a \leq 2.50$), middle ($2.50 < a \leq 2.82$),
and outer ($2.82 < a \leq 3.27$) regions, where $a$ is the semimajor axis in AUs. 
It should be noted that some near-Earth asteroids (Apollos and Amors) 
which occasionally have semimajor axis in the ranges of MBAs, are not considered 
in the following discussion, and the other group of near-Earth asteroids in which $a<1$ AU (Atens) are, by definition, 
not within main belt regions. 

\begin{figure*}
\epsscale{1.} \plotone{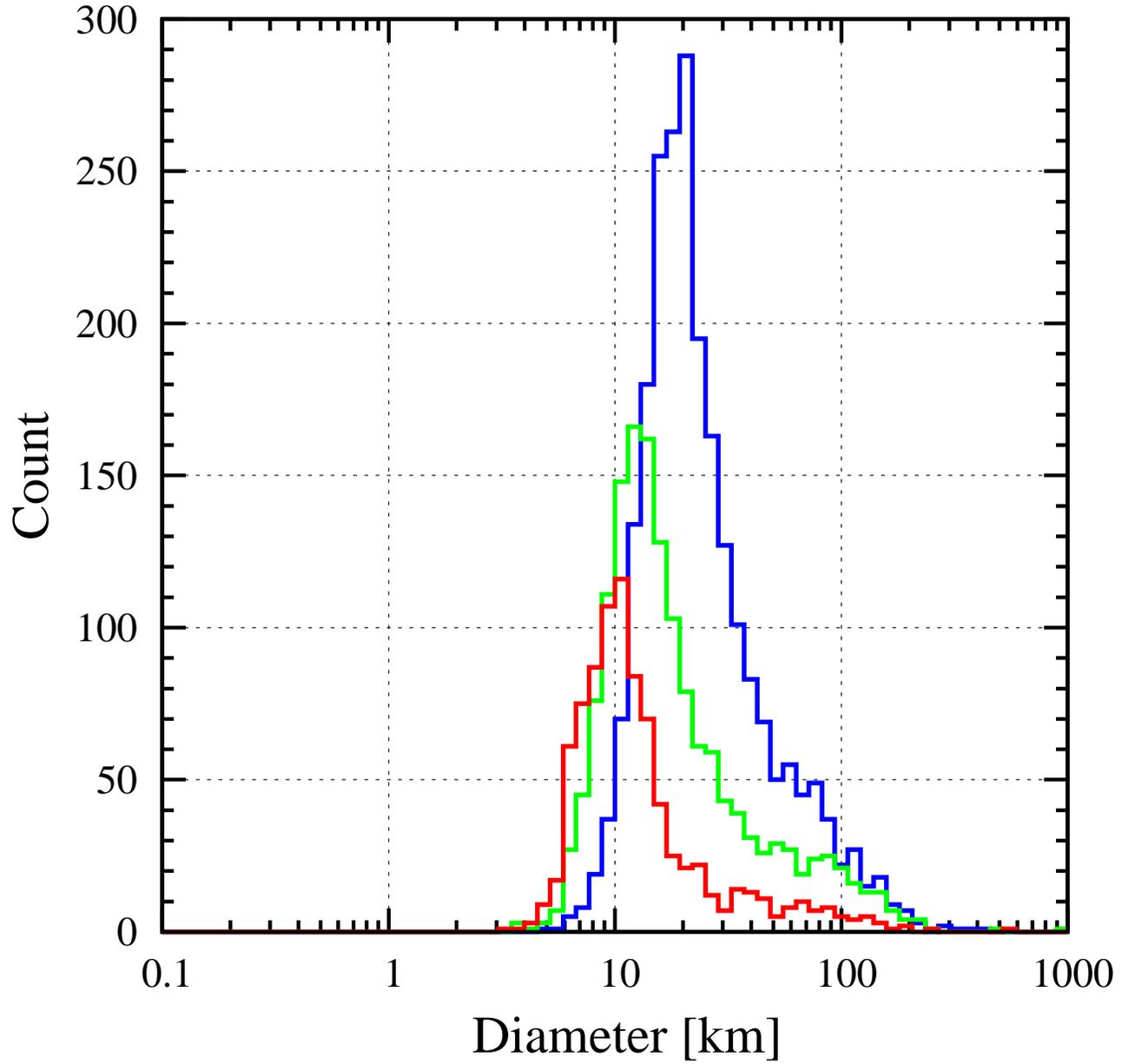}
\caption{Histogram of the size distribution of AcuA MBAs. 
Red, green, and blue lines denote 
inner, middle, and outer region MBAs, respectively. }
\label{number count MBA}
\end{figure*}

The orbital elements of the asteroids were obtained from the Asteroid Orbital Elements Database ~\citep{Bowell1994} 
distributed by Lowell Observatory\footnote{The data are available at ftp://ftp.lowell.edu/pub/elgb/astorb.html}.
The boundaries of the main belt regions at the semimajor axis $a=2.06$, $2.50$, $2.82$, and $3.27$ AU correspond, 
respectively, to the 4:1, 3:1, 5:2, and 2:1 mean motion resonances of Jupiter 
(e.g., ~\citet{Kirkwood1867, Froeschle1989, Scholl1989, Yoshikawa1989}).
The mean motions and secular resonances of Jupiter are the dominant effects on the dynamical evolution of MBAs
at the present stage of solar system evolution. 
The Yarkovsky thermal force ~\citep{Bottke2002b} is also known to cause changes in 
the orbits of asteroids (mainly affecting the semimajor axis), 
but its effects are less than those of the dynamical resonances on large-size objects. 
The numbers of AcuA asteroids in each region of the main belt (inner, middle, and outer) 
is 858, 1523, and 2341, respectively 
(also see Table \ref{number of asteroids}). 
The number of detected asteroids in the outer region is greater than that in the inner region, implying that 
the actual number of asteroids is greater in outer regions, compared to the number of observed 
asteroids, which generally decreases 
as a function of distance on account of distance-dependent instrumental detection limits; however, 
the precise number density of asteroids in each region was not examined. 
Moreover, the number of inner region MBAs is about half that of middle region MBAs, despite the fact that 
inner region MBAs are more readily detected.
Figure \ref{number count MBA} also illustrates the size distribution of the inner, middle, and outer MBAs. 
Concerning 
the size distribution of asteroids, the number of asteroids is expected to increase monotonically with decrease in size. 
This figure, however, shows maxima at $d\sim20$ km for the outer MBAs. 
This suggests that the survey completeness rapidly drops for asteroids 
smaller than $\sim20$ km in the outer region on account of the detection limit of {\it AKARI} instrumentation. 
This size limit is consistent with the fact that 
AcuA is complete for all MBAs with $H < 10.3$, as described in Section \ref{AcuA data}.
In this paper, we consider only the inner, middle, and outer MBAs in our comprehensive 
analysis of the asteroid belt; we do not consider objects outside of
the main belt ($ a \leq 2.06$, or $a > 3.27$)
and do not investigate the properties of the individual dynamical families in detail, which will be 
discussed in separate papers; e.g., the Cybele family is studied in~\citet{Kasuga2012}.

\section{Albedo properties of MBAs}
\subsection{Albedo size-dependencies}
\begin{figure*}
\begin{center}
\hspace*{-1cm}
\begin{tabular}{cccc}
\hspace{1cm}{\small (a) MBAs total} &%
\hspace{1cm}{\small (b) inner MBAs} \\
\resizebox{85mm}{!}{\epsscale{2.} \plotone{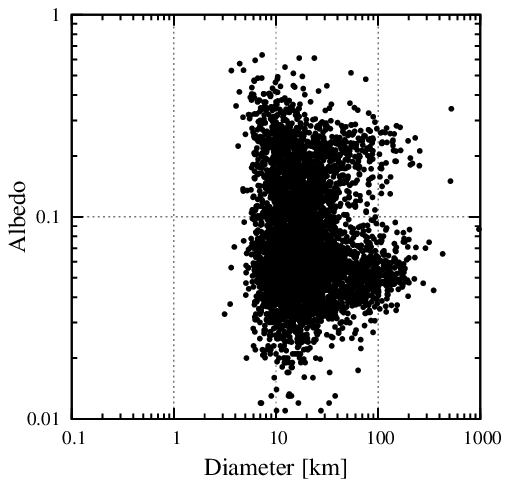}}&%
\resizebox{85mm}{!}{\epsscale{2.} \plotone{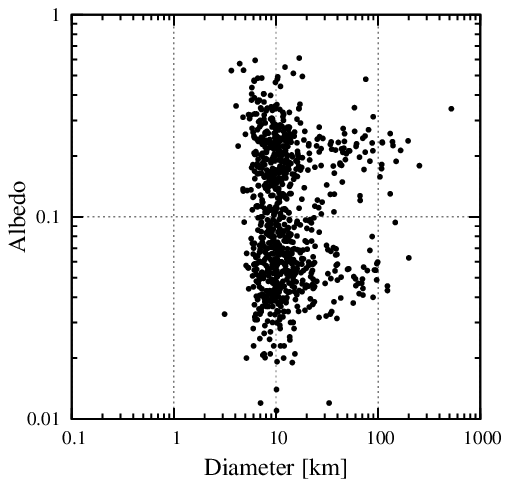}}\\
\\
\hspace{1cm}{\small (c) middle MBAs} &%
\hspace{1cm}{\small (d) outer MBAs} \\
\resizebox{85mm}{!}{\epsscale{2.} \plotone{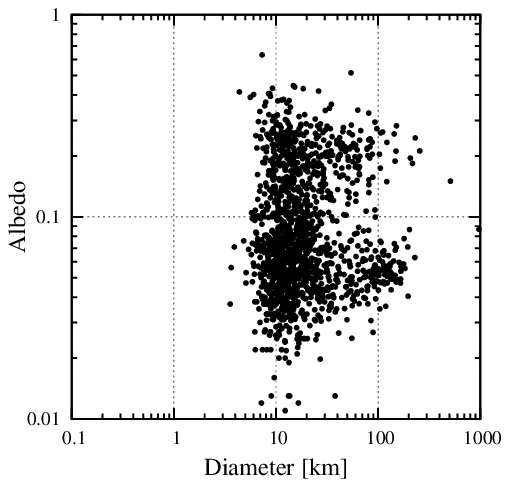}}&%
\resizebox{85mm}{!}{\epsscale{2.} \plotone{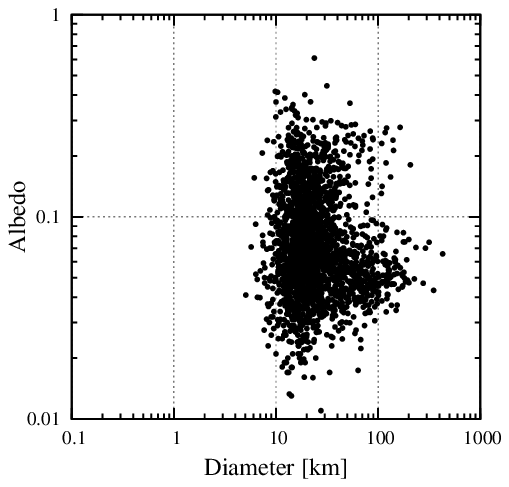}}\\
\end{tabular}
\end{center}
\caption{Size--albedo distributions for AcuA MBAs: (a) total population, and (b) inner, (c) middle, 
and (d) outer region MBAs. %
} \label{MBA size vs albedo}
\end{figure*}

Figure \ref{MBA size vs albedo} shows the distributions of albedo values as a function of asteroid diameter ($d$),
for asteroids cataloged in the AcuA. 
The total number of asteroids with $d<5$ km is relatively low, on account of the detection limit for small bodies 
(see ~\citet{Usui2011}, Figure 8a). 
Notably, the distributions of albedo values for the asteroids in each region are bimodal. 
	
\begin{figure*}
\hspace*{-1cm}
\begin{tabular}{cccc}
\hspace{1cm}{\small (a) C-type} &%
\hspace{1cm}{\small (b) S-type} \\
\resizebox{85mm}{!}{\epsscale{1.} \plotone{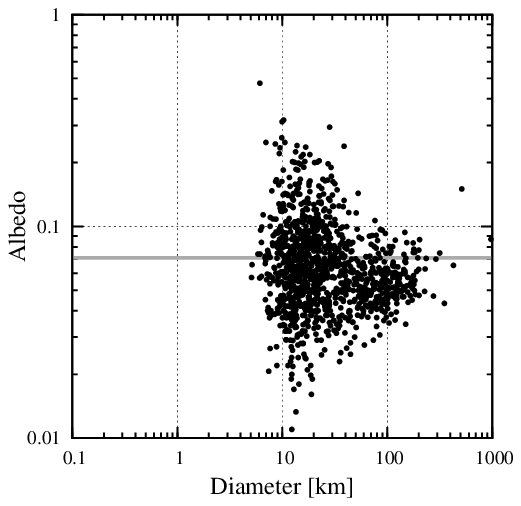}}&%
\resizebox{85mm}{!}{\epsscale{1.} \plotone{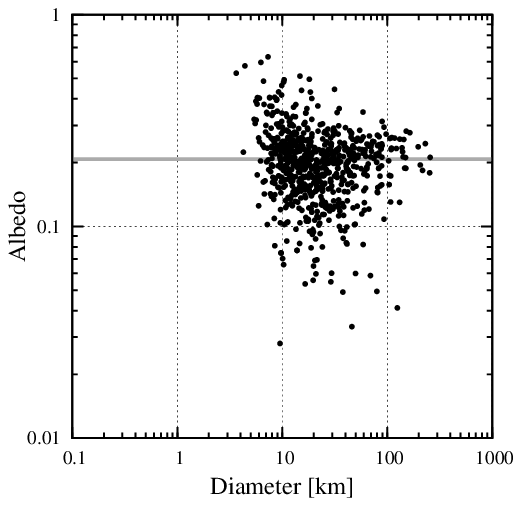}}\\
\\
\hspace{1cm}{\small (c) X-type} &%
\hspace{1cm}{\small (d) D-type} \\
\resizebox{85mm}{!}{\epsscale{1.} \plotone{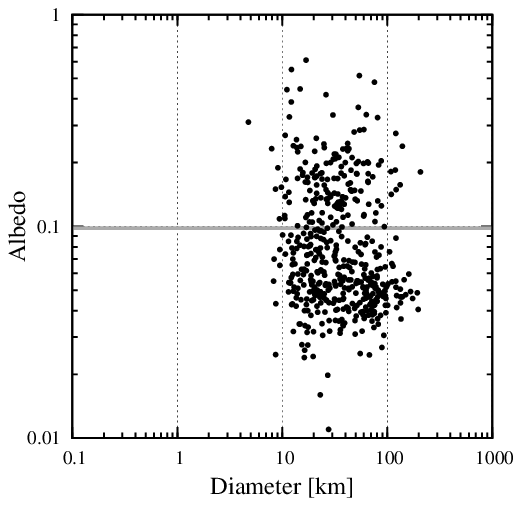}}&%
\resizebox{85mm}{!}{\epsscale{1.} \plotone{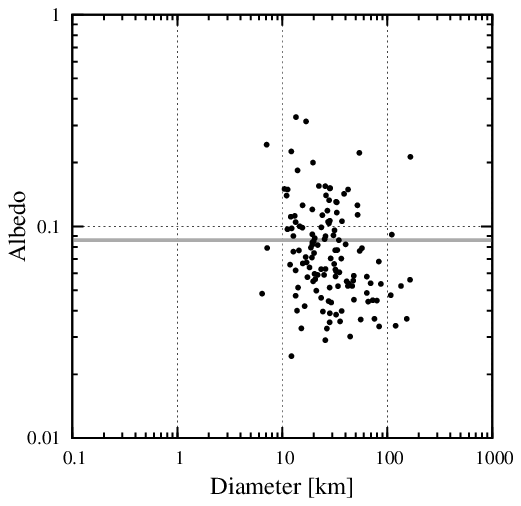}}\\
\end{tabular}
\caption{
Size--albedo distributions for each taxonomic type of AcuA MBAs. 
Solid gray lines indicate the mean albedo values (see Table \ref{table:summary for 5 types}). 
}
\label{fig:summary for taxonomic classes}
\end{figure*}

Figure \ref{fig:summary for taxonomic classes} shows the distribution of albedo values of 
AcuA MBAs as a function of asteroid size, 
separated by taxonomic type. It should be noted that the taxonomic type is known for 
2496 AcuA asteroids (53\% of the total number of detected MBAs). 
It should also be noted that the taxonomic type was determined using data from the literature mentioned above, 
and not from {\it AKARI} observations; this is because (1) the {\it AKARI} observations were conducted at 
infrared, not optical, wavelengths, and (2) the two {\it AKARI} mid-infrared 
channels ({\it S9W} and {\it L18W}) 
did not observe the same region of the sky simultaneously; thus, 
a complete data set for each asteroid was not always obtained (see ~\citet{Usui2011}, details).
Figure \ref{fig:summary for taxonomic classes} reveals that 
the clusters of high- and low-albedo asteroids depicted in Figure \ref{MBA size vs albedo} 
correspond to S-type and C-types asteroids. 
In both distributions, it is found that the scatter in albedo among the smaller asteroids 
increases. The albedo values of X-type asteroids are widely distributed. The albedo values of D-type 
asteroids are moderately low, and the number of D-type asteroids is
relatively small (see Table \ref{table:summary for 5 types}). 

\begin{figure*}
\begin{center}
\resizebox{85mm}{!}{\epsscale{1.0} \plotone{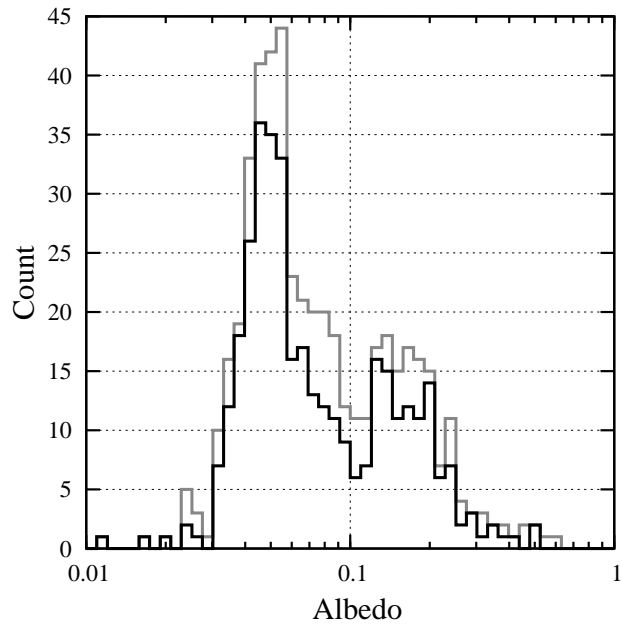}}
\end{center}
\caption{
Histogram of albedo values for AcuA X-type MBAs. 
Gray and black lines denote all X-types, and X-types larger than 20km, respectively. 
The bin size is set at 50 segments for the albedo range of 0.01--1.0 
in the logarithmic scale.
}
\label{fig: MBA X-type albedo histogram}
\end{figure*}

X-type asteroids, which have featureless flat or slightly 
reddened spectra over optical wavelengths, are spectrally degenerate  
and can be classified into three subclasses only on the basis of albedo values 
~\citep{Tholen1989, Clark2004, Fornasier2011}, as 
E-type: $p_{\rm v} > 0.3$ (high albedo), M-type: $0.3 \ge p_{\rm v} > 0.1$ (medium albedo), and
P-type: $0.1 \ge p_{\rm v}$ (low albedo). 
These three classes are spectrally similar to each other, but have very different inferred mineralogy, as 
E-type: containing enstatite-rich aubrites, M-type: containing metallic iron cores, and
P-type: carbon and/or organic-rich. According to the classification based on albedo values, of the 490 X-type AcuA MBAs, 
14 are E-type, 145 are M-type, and 331 are P-type. Note that 
for 90\% of these X-type MBAs, the AcuA albedos provide the first sub-classification into E-, M-, and P-types.
The largest E-type member is (71) Niobe ($d=80.86$ km, $p_{\rm v} = 0.326$) in the middle MBAs.
The mean albedos of each subclass are presented in Table \ref{mean albedo for X-type}. 
Note that there are two objects that show inconsistency  between taxonomic type and albedo value. 
While (498) Tokio was classified as an 
M-type asteroid in a previous work ~\citep{Tholen1984}, the albedo value of this object
is $p_{\rm v} = 0.063$ by {\it AKARI} (or $0.069$ by the {\it IRAS}); thus, (498) is a 
P-type, not an M-type, asteroid. (55) Pandora was also classified as an 
M-type asteroid (also by ~\citet{Tholen1984}); however, it is an 
E-type asteroid ($p_{\rm v} = 0.337$ by {\it AKARI} or $0.301$ by the {\it IRAS}). 
(498) and (55) are middle MBAs.
Figure \ref{fig: MBA X-type albedo histogram} shows the albedo distribution of X-type asteroids. 
There are two major components found in this figure, 
the M-type with $p_{\rm v} > 0.1$ and P-type with  $p_{\rm v} \le 0.1$. 
From this distribution, it seems reasonable that the boundary between medium-albedo M-type 
and low-albedo P-type is set as $p_{\rm v} = 0.1$. On the other hand, no clear boundary is found 
around  $p_{\rm v} \sim 0.3$, between E- and M-type. 

Two large and high-albedo asteroids are observed among the D-types:  
(9) Metis ($d=166$ km, $p_{\rm v}= 0.213$) is an inner MBA, 
and (224) Oceana ($d=54$ km, $p_{\rm v}= 0.222$) is a middle MBA.
These were considered as D-type asteroids by ~\citet{Lazzaro2004}, but were recently 
reclassified as S-type (9) and M-type (224) asteroids ~\citep{Neese2010};
albedo values for these asteroids in the AcuA also support this classification of ~\citet{Neese2010}.

\subsection{Variations in the distributions of albedo values}

\begin{figure*}
\hspace*{-1cm}
\begin{tabular}{ccc}
\hspace{1cm}{\small (a) $0.1 \leq d < 20$ km}  &%
\hspace{1cm}{\small (b) $20 \leq d < 100$ km}\\
\resizebox{85mm}{!}{\epsscale{1.0} \plotone{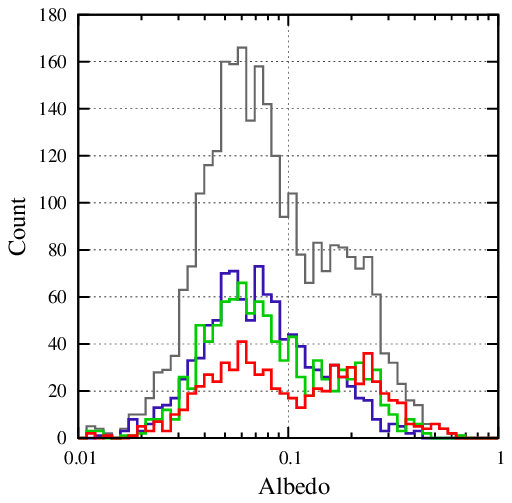}} &%
\resizebox{85mm}{!}{\epsscale{1.0} \plotone{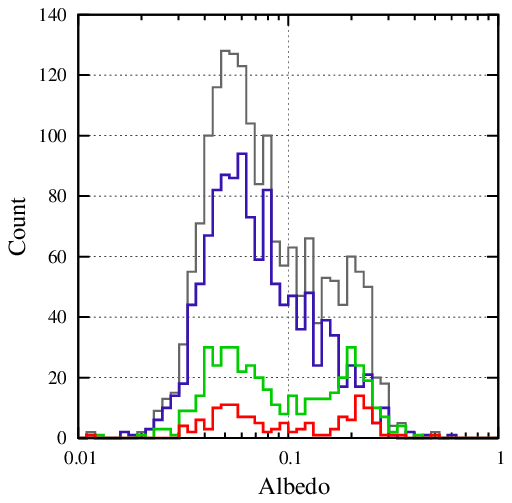}}\\
\\
\hspace{1cm}{\small (c) $d \geq 100$ km}  \\
\resizebox{85mm}{!}{\epsscale{1.0} \plotone{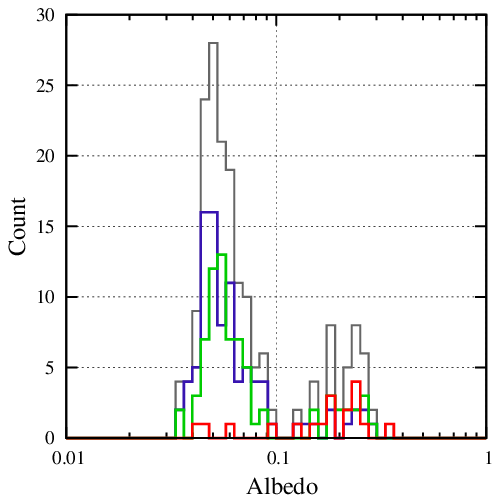}} \\
\end{tabular}
\caption{
Histogram of albedo values for MBAs detected by {\it AKARI} for three size classes 
(diameters, $d$, in km): (a) $0.1 \leq d < 20$; (b) $20 \leq d < 100$; (c) $d \geq 100$.
Red, green, blue, and gray lines denote 
inner region, middle region, outer region, and total MBAs, respectively. 
The bin size is set at 50 segments for the albedo range of 0.01--1.0 
in the logarithmic scale.
}
\label{fig: MBA inner-middle-outer}
\end{figure*}

Figure \ref{fig: MBA inner-middle-outer} shows histograms of asteroid albedo values for 
the inner, middle, and outer region. 
In Figure \ref{MBA size vs albedo} (also see ~\citet{Usui2011}, Figure 8b), 
the ``bimodal'' distribution of albedo is clearly visible. 
The population densities of the low- and high-albedo components are, however, not the same; 
the peak of the low albedo distribution is about twice that of the high albedo distribution. 
This same pattern is observed 
in the inner, middle, and outer region distributions as Figure \ref{MBA size vs albedo}, 
but the detailed features of each distribution vary. 
In the inner region (red lines in Figure \ref{fig: MBA inner-middle-outer}), 
the peak of the high albedo distribution is nearly the same as, or a little higher than, 
that of the lower albedo distribution. 
In the middle and outer region asteroids (green and blue lines in Figure \ref{fig: MBA inner-middle-outer}), 
the low albedo component is dominant, and 
the peak of the high albedo distribution is nearly buried in the long tail of 
the low albedo component; this pattern is especially prominent 
in the albedo distributions of the smaller asteroids. 

Another outstanding feature of the albedo distributions is that,
for the large asteroids (Figure \ref{fig: MBA inner-middle-outer}c), 
the distributions are clearly divided at $p_{\rm v} \sim 0.1$; 
this division is less pronounced in the small asteroids; 
the mean albedo value of the total population of 4722 AcuA MBAs is $\overline{p_{\rm v}} = 0.102 \pm 0.079$. 
This division in the albedo distribution 
was also observed in data from the {\it IRAS} asteroid catalog ~\citep{Tedesco2002};
however, the boundary was observed at $p_{\rm v} = 0.089$ ~\citep{Morbidelli2002}. 
Similar features are found in the {\it WISE}/NEOWISE data set ~\citep{Masiero2011}. 

\begin{figure*}
\begin{center}
\begin{tabular}{c}
(a)\\
\resizebox{95mm}{!}{\epsscale{1.0} \plotone{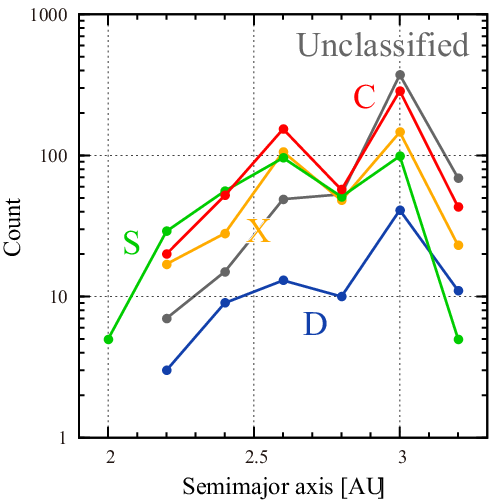}}\\
\\
(b)\\
\resizebox{95mm}{!}{\epsscale{1.0} \plotone{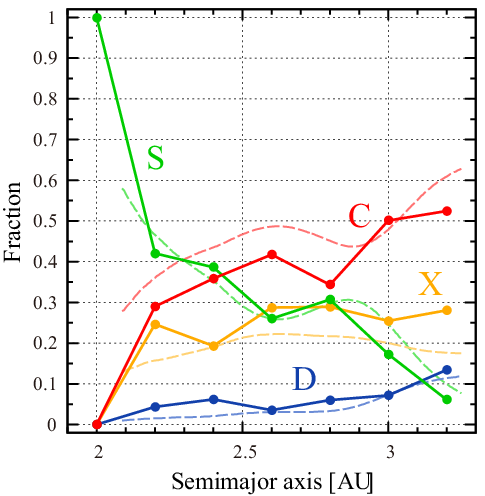}}\\
\end{tabular}
\end{center}
\caption{%
Histograms showing the numerical distribution (a) and fractional distribution (b) of MBAs 
with $d \ge 20$ km  for each of the taxonomic types.
Bold lines in (b) show AcuA data (this study) and dashed lines show data from \citet{Bus2002}. 
Red, green, yellow, and blue denote C-, S-, X-, and D-type asteroids, respectively.}
\label{fig: MBA taxonomy fraction}
\end{figure*}

Figure \ref{fig: MBA taxonomy fraction} shows 
the heliocentric distribution for asteroids with diameters $d > 20$ km for each of the taxonomic types. 
AcuA covers a complete data set of all MBAs with $d > 20$ km. 
We should note that there is a possible selection effect in the determination of the taxonomic type, 
i.e., spectroscopic data have been preferentially obtained from 
brighter (larger size, higher albedo) objects 
and/or particular family members, and hence these objects are more heavily represented 
in the data than are darker objects.

Figure \ref{fig: MBA taxonomy fraction}a illustrates the numbers of AcuA MBAs of each taxonomic type. 
As observed in Figure \ref{semj count}, the number of detected asteroids increases with increasing 
semimajor axis, except for large deviations at $a\sim2.8$ AU. 
Taxonomic classifications were determined for 1409 asteroids 
(71\% of the AcuA MBAs with sizes $d > 20$ km), 
representing the following regional distribution: 144 inner, 463 middle, and 802 outer MBAs. 
Percentage of asteroids with taxonomy determinations in each region is 92\%, 88\%, and 62\%, respectively. 
Fewer classifications exist in outer region  
due to the selection effect as mentioned above. Figure \ref{fig: MBA taxonomy fraction}b shows 
the fractional representation of each taxonomic type at different semimajor axis, as well 
as the results obtained by \citet{Bus2002} (dashed lines). 
\citet{Bus2002} showed the bias-corrected heliocentric distribution 
for each taxonomic type in 1447 asteroids of size $d > 20$ km, which is the same size range in this work. 
We observed distributions of S-type and D-type asteroids 
similar to those of \citet{Bus2002}; that is, S-type asteroids are dominant in the inner region, 
with relative proportions decreasing with an increase in semimajor axis; 
D-type asteroids comprise $< 10$\% of the total number of asteroids, 
but their relative proportions gradually increase with increasing semimajor axis.  
Our data on the fractions of C-type and X-type asteroids depart from those of \citet{Bus2002} by $\sim10$\%; 
the fraction of C-type in this work is less than that in \citet{Bus2002}, and 
that of X-type is more than that in \citet{Bus2002}. 
This result could be caused by biases related to the number of taxonomic identifications. 
In the literature, X-types crowd around 3 AU \citep{Mothe-Diniz2003} and 
the Hungaria family~\citep{Warner2009}. The former group can be confirmed from our sample 
in Figure \ref{fig: MBA taxonomy fraction}b. 
The latter is in the range of $1.78 < a < 2.06$ AU~\citep{Gradie1979} and out of the MBAs. 
The number of AcuA V-type asteroids is small 
((4) Vesta, (854) Frostia, (1273) Helma, (1981) Midas, and (3657) Ermolova; 
(1981) is a near-Earth asteroid (Apollos), not a MBA, and the other four are inner MBAs), 
and (4) is the only asteroid larger than 20 km. These V-type asteroids are excluded from this study.

\begin{figure*}
\begin{center}
\begin{tabular}{c}
(a)\\
\resizebox{95mm}{!}{\epsscale{1.0} \plotone{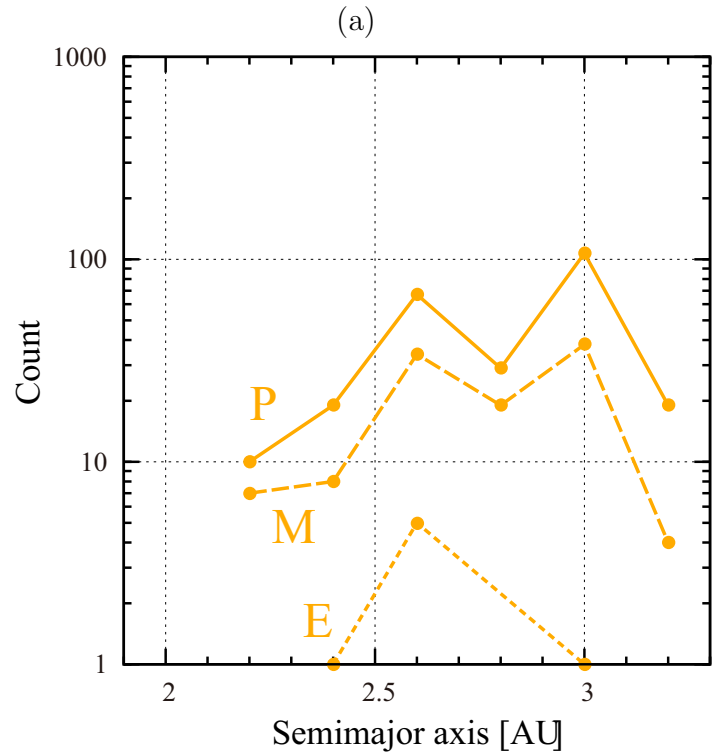}}\\
\\
(b)\\
\resizebox{95mm}{!}{\epsscale{1.0} \plotone{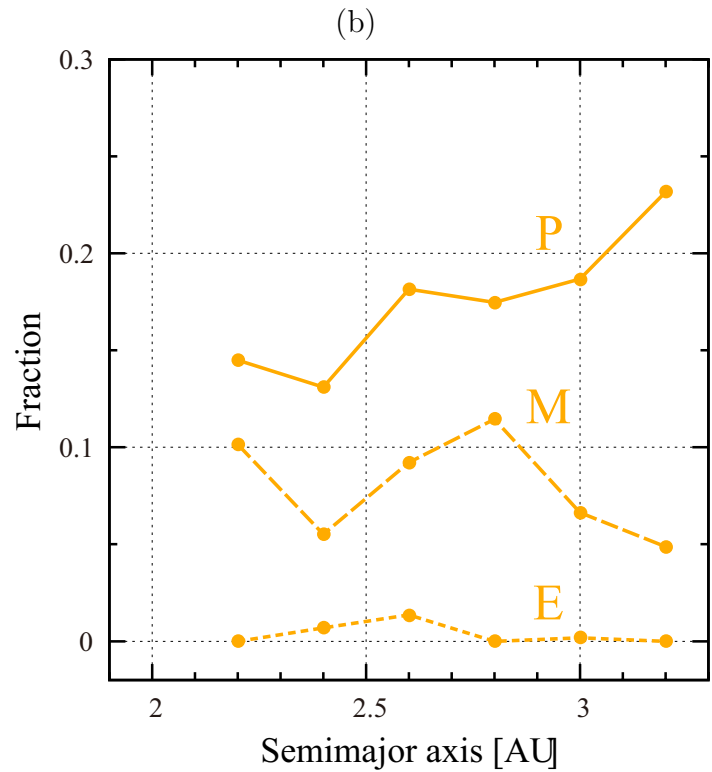}}\\
\end{tabular}
\end{center}
\caption{%
Histograms showing the numerical distribution (a) and fractional distribution (b) of X-type MBAs 
with $d \ge 20$ km. 
Dotted, dashed, and bold lines denote E-, M-, and P-type asteroids, respectively.}
\label{fig: MBA X-type taxonomy fraction}
\end{figure*}

Figure \ref{fig: MBA X-type taxonomy fraction} 
shows the heliocentric distribution for subclasses of X-types from AcuA asteroids 
with diameters $d > 20$ km: E-, M-, and P-type. 
Traditionally the distribution of X-subclasses were described in ~\citet{Bell1989} 
based on Tholen taxonomy~\citep{Tholen1984}, however they did not have detailed albedo information. 
Figure \ref{fig: MBA X-type taxonomy fraction} 
shows revised distributions based on the AcuA data set. The number of identified E-type asteroids 
larger than 20 km is only seven (see, Table \ref{E-class asteroids}). 
These large E-type asteroids are non-family members except for (44) Nysa. Although 
E-type members are primarily located in the inner asteroid belt, five of seven large E-types 
are the middle MBAs. 
One E-type in the outer MBAs is (665) Sabine ($d=53.01$ km, $p_{\rm v} = 0.365$, $a=3.14$ AU), 
which has a detailed shape model~\citep{Michalowski2006}. 
A major group in the X-subclasses is the P-type, which accounts for 
68\% of our X-type sample ($d > 20$ km). 
Throughout the main belt region, P-type asteroids are dominant among the X-types. 
In Figure \ref{fig: MBA X-type taxonomy fraction}b, 
the abundance of P-types increases beyond 3 AU, while that of M-types decreases. 
The distribution of P-types is similar to that of C- or D-types 
in Figure \ref{fig: MBA taxonomy fraction}b. The mean albedo of P-types 
is $0.054 \pm 0.017$ of total 251 (Table \ref{mean albedo for X-type}), 
which is also similar to C: $0.066 \pm 0.031$, 
or D: $0.077 \pm 0.041$  (Table \ref{table:summary for 5 types}).
From this, it can be conjectured that P-type asteroids have similar origin or similar evolutional process, 
to C- or D-types, though the actual component of X-types are not fully 
understood yet. 

\begin{figure*}
\hspace*{1cm}
\begin{tabular}{cccc}
\hspace{1.5cm}{\small (a) total}\\
\resizebox{65mm}{!}{\epsscale{1.0} \plotone{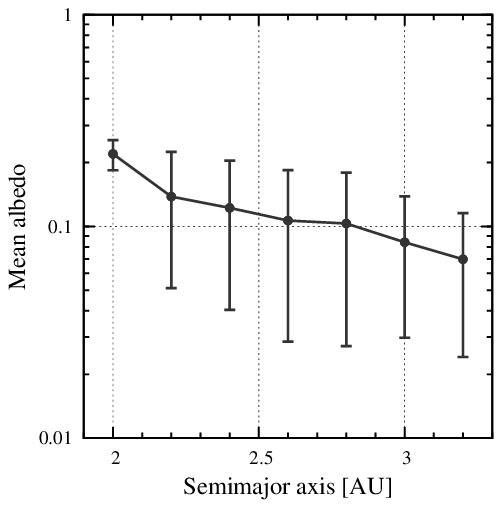}}\\
\\
\hspace{1.5cm}{\small (b) C-type} &
\hspace{1.5cm}{\small (c) S-type} \\
\resizebox{65mm}{!}{\epsscale{1.0} \plotone{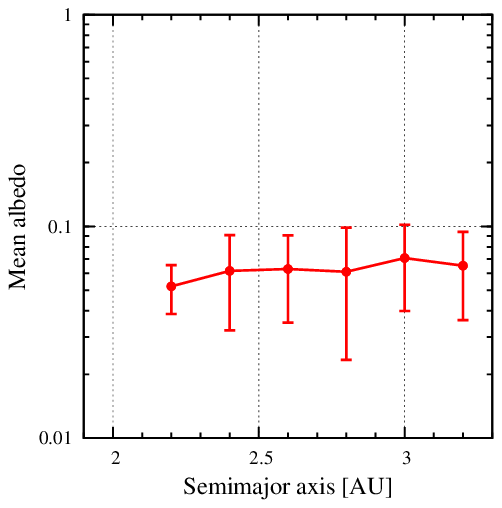}} &
\resizebox{65mm}{!}{\epsscale{1.0} \plotone{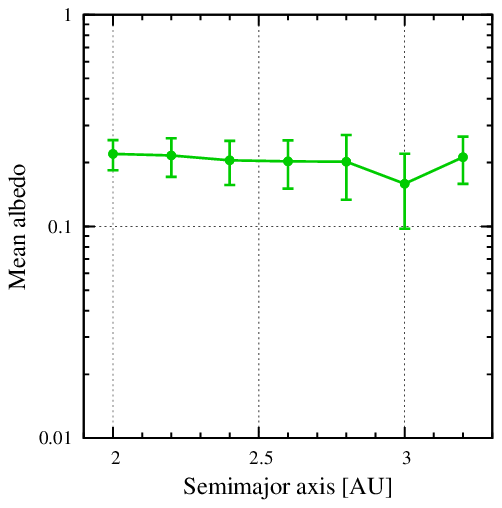}} \\
\\
\hspace{1.5cm}{\small (d) X-type} &
\hspace{1.5cm}{\small (e) D-type}  \\
\resizebox{65mm}{!}{\epsscale{1.0} \plotone{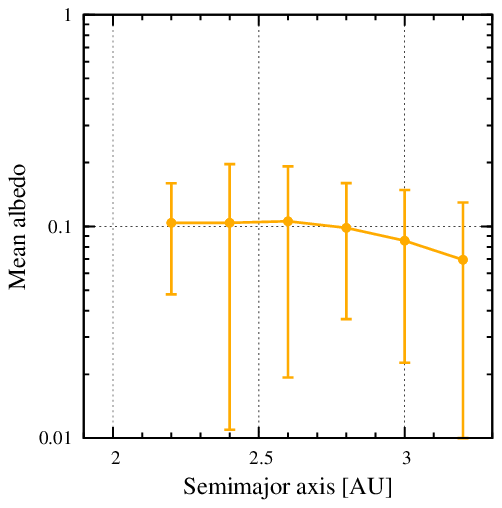}} &
\resizebox{65mm}{!}{\epsscale{1.0} \plotone{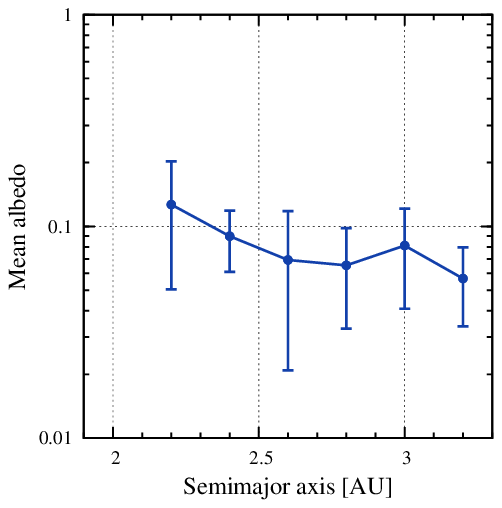}} \\
\\
\end{tabular}
\caption{%
Mean albedo as a function of the semimajor axis for MBAs with $d \ge 20$ km for each of the taxonomic types. 
Bars represent variations of albedo in each region. }
\label{fig: MBA inner-middle-outer type stacked ratio 20km}
\end{figure*}

Figure \ref{fig: MBA inner-middle-outer type stacked ratio 20km} presents the dependency of mean albedo on 
heliocentric distance. In the total population of asteroids, albedo gradually decreases as the 
semimajor axis increases; i.e., asteroids located further from the sun are darker. 
On the other hand, within individual taxonomic types, 
the heliocentric distributions of mean albedo are nearly constant throughout the entire main belt, 
although albedo variations at any given distance are large. 
Thus, the albedo within a given taxonomic type is relatively independent of heliocentric distance. 
The cause of the decline in the mean albedo of the total distribution in outer regions is not 
the heliocentric distance itself, but
the influence of the compositional mixing ratios of taxonomic types, as observed 
in Figure \ref{fig: MBA taxonomy fraction}. 

\section{Discussion}

As observed in Figure \ref{fig:summary for taxonomic classes}, the albedo of asteroids is dependent on its size.
This does not mean, however, that smaller asteroids tend to have higher albedos; 
rather, it implies that the diversity of albedo values is greater for smaller asteroids than for larger asteroids. 
AcuA provides a complete data set for $H < 10.3$, which corresponds to $d > 20$ km. 
Moreover, detectability with {\it AKARI} does not depend on albedo. This is because 
the thermal flux of an asteroid is hardly dependent on albedo 
(surface temperature is proportional to $(1-A_{\rm B})^{1/4}$, which has a 
value in a range from 0.99 to 0.93 with $p_{\rm v}$ of 0.01 to 0.6, 
where $A_{\rm B} = (0.290 + 0.684 G)~p_{\rm v}$ is the Bond albedo), 
while the visible reflected component of sunlight is proportional to the albedo. 
For these reasons, the scatter in albedo among the smaller objects should be real,
not a bias effect; even so, 
taxonomic information is available only for 53\% of the total MBAs in AcuA, 
or 71\% of $d > 20$ km MBAs in AcuA. 
Taxonomic classifications are derived from spectroscopic studies of asteroids, 
which have strong observational selection biases. Especially for asteroid with $d < 20$ km, 
our study of the size-albedo distribution of MBAs rapidly becomes incomplete due to the sensitivity limits of {\it AKARI}.

Spacecraft explorations show that most asteroid surfaces are 
covered with regolith and/or numerous boulders in a wide range of sizes. 
For example, a unique boulder with an unusually low brightness is found to sit on top of the asteroid (25143) Itokawa 
like a benchmark ~\citep{Fujiwara2006}.
Several interpretations for the origin of this distinct ``black boulder'', which is about 40\% darker than 
surrounding materials, have been considered (e.g., \citet{Hirata2011}). 
Such local heterogeneities are averaged out and nearly homogenized in the global view, and 
only a small degree of regional variation on surfaces is observed. 
However, especially for smaller asteroids, the effects of these heterogeneities are conspicuous 
relative to their sizes. This could partly explain the albedo variations at smaller sizes.

For S-type asteroids, albedo dependency can be explained by 
the space weathering of stony chondritic asteroids. 
The collisional life times and surface ages of S-type asteroids are correlated with their sizes 
~\citep{Davis2002, Bottke2005, Nesvorny2005}; thus, fresher  and smaller objects 
are less space weathered and their albedos are relatively higher ~\citep{Burbine2008, Thomas2011}. 
Space weathering effects cause the visible and near-infrared reflectance spectra of stony asteroids to be
darker and redder than those of pristine materials ~\citep{Sasaki2001}. 
Ground-based observations suggest that a color 
transition exists between ordinary chondrite-like objects and S-type 
asteroids over the size range 0.1--5 km, which implies that the surfaces 
of small stony asteroids are evolving towards the colors of large 
asteroids ~\citep{Binzel2004, Binzel2010}. Our data indicate that the albedo transition 
occurs in the size range of $d > 5$ km. 

\begin{figure*}
\epsscale{1.} \plotone{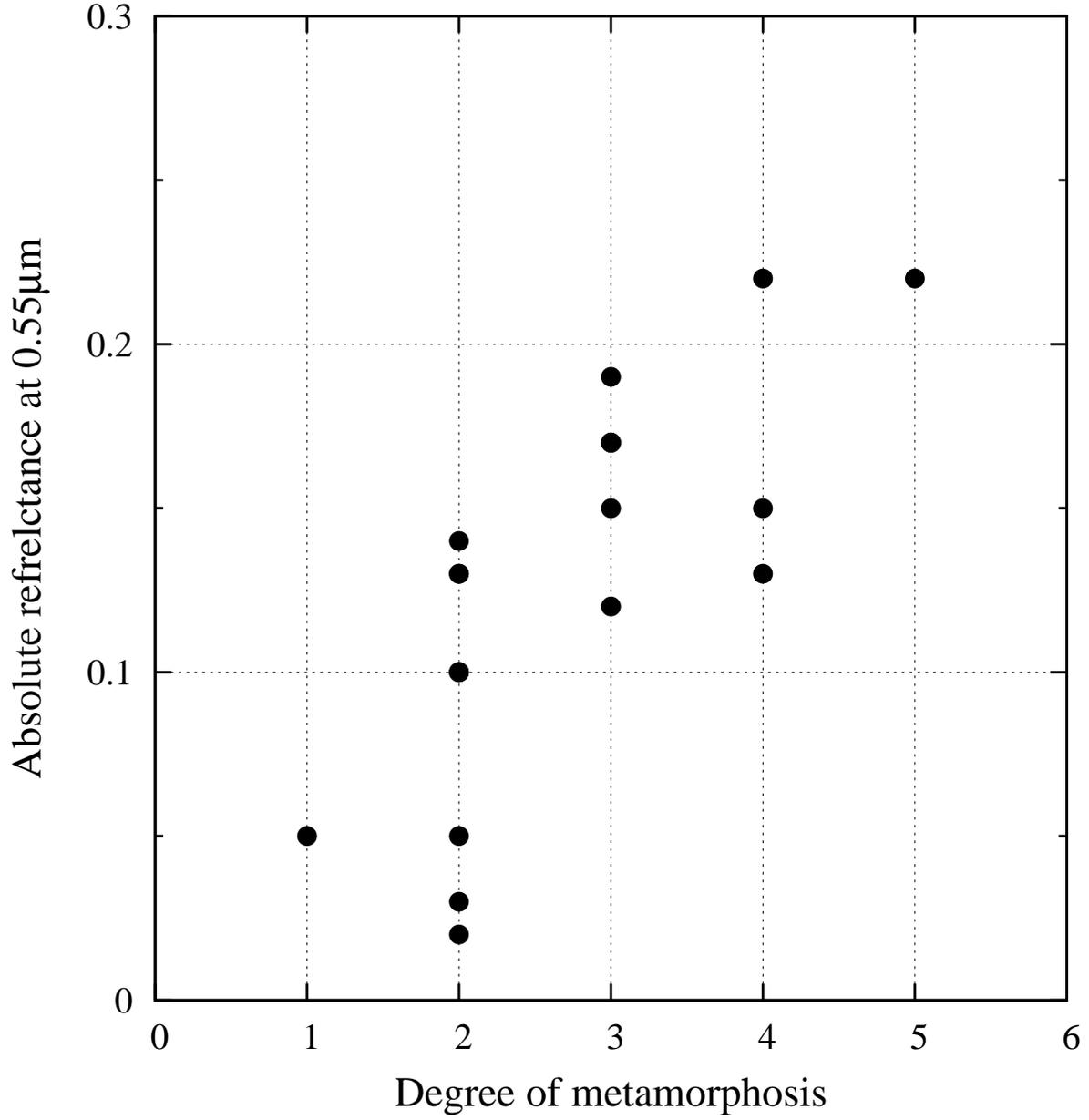}
\caption{Absolute reflectance at 0.55 \micron\ as a function of the degree of metamorphosis/alteration
of meteorites measured in laboratories; compiled by \citet{Clark2009}.}
\label{thermally metamorphosis}
\end{figure*}

In contrast to S-type asteroids, the reason for the albedo transition in C-type asteroids is less known. 
The relationship between age and the space weathering effect in C-type asteroids 
has been mentioned in several papers (e.g., ~\citet{Nesvorny2005, Lazzarin2006}),  however, 
the relationship is still uncertain, mainly on account of the ``darkness'' of C-type asteroids. 
Laboratory measurements indicate that 
thermally metamorphosed samples of carbonaceous chondrites possess high absolute reflectance ~\citep{Hiroi1994}, 
which could cause the high albedo of asteroids. Figure \ref{thermally metamorphosis} shows 
the dependence of the absolute reflectance of chondrites on the degree of 
metamorphism~\citep{Clark2009}. The classification of metamorphism in chondritic meteorites 
is qualitatively determined ~\citep{Weisberg2006}:
the most pristine materials are type 3; types 4 to 6 represent an increasing degree of thermal
metamorphism, and types 2 to 1 represent an increasing degree of aqueous alteration. While 
thermal metamorphism and aqueous alteration are related to different mechanisms of recrystallization, 
the classification from 1 to 6 can be taken to represent an increasing degree of metamorphism. 
The metamorphism of chondrites occurs in the inner parts of the parent body, which is internally 
heated by the decay energy of short-lived radioisotopes (e.g., \citet{Trieloff2003, Greenwood2010}). 
In a catastrophic disruption caused by a highly energetic impact, 
small fragments, including those from the inner portion of a parent body, are exposed. 
Asteroids smaller than a few 10s of kilometers are considered to be formed mainly 
by collisional breakup ~\citep{Bottke2005}; thus, the high albedo of some materials may have 
been derived from deeper thermal metamorphic processes. 
In other materials, however, the albedo might be due to the presence of reaccumulated fragments 
(e.g., ~\citet{Davis1979, Fujiwara1982}), 
and/or a regolith layer ~\citep{Hiroi1992, Hiroi1994a}. 
Thus, a variety of processes might cause the albedo variation at smaller sizes. 

No clear dependency of albedo on size is found for D-type asteroids in Figure \ref{fig:summary for taxonomic classes}d, 
while the number of detected D-types is relatively small. 
Some researchers have distinguished between the features of inner D-type and outer D-type varieties of 
D-type asteroids (e.g., \cite{Lagerkvist2005, Mothe-Diniz2009}). 
D-type asteroids dominate the Jovian Trojans, but become rarer at smaller heliocentric distances. 
As discussed by \citet{Carvano2003}, the variations in D-type asteroids might be related 
to different origins and evolutionary histories, or 
to different processes of differentiation occurring closer to the sun.
Indeed, the actual mineralogical characterization of D-type asteroids is difficult to ascertain 
because of their small number of samples (except for the Tagish Lake meteorite; ~\citet{Hiroi2001}). 

\section{Summary}

We have presented an analysis of the albedo properties of MBAs detected by {\it AKARI} and recorded in the AcuA.
Of the 5120 AcuA asteroids, 4722 are MBAs, 
and 1974 are provided as a complete data set of MBAs larger than $d > 20$ km. 
As is already known, 
the albedo distribution of MBAs is strongly bimodal; this trend is present not only in the distribution of 
the total population, but also in the distributions of inner, middle, and outer  MBAs. The bimodal distributions 
are separated into two major 
groups at an albedo value of $p_{\rm v} = 0.1$, which demarcates low-albedo C-type and high-albedo S-type asteroids. 	
In each group, the albedo distribution is size-dependent, 
and the variation in albedo values is greater at smaller sizes. 
For smaller asteroids, the effects of surface heterogeneities 
on albedo are relatively large, while
such local heterogeneities are averaged out and seemingly homogenized for larger asteroids.
Moreover, albedo distributions in S-type asteroids appear to be affected 
by the space weathering, whereas the albedo distributions in C-type asteroids are 
partially explained by the effects of metamorphism.

We examined the heliocentric distributions of mean albedo values for each taxonomic type. 
In spite of the influence of the space weathering and other albedo transition processes, 
the mean albedo is nearly constant and independent of heliocentric distance throughout 
the entire main belt region, irrespective
of taxonomic type. In the total distribution, on the other hand, 
the mean albedo value gradually decreases with increasing the semimajor axis, 
presumably due to the compositional mixing ratios of taxonomic types. 

Almost 90\% of the X-type MBAs in the AcuA data set can now be subdivided into the E-, M-, and P-types 
based on the {\it AKARI}-derived albedos. 
The distribution of P-types, which have lower albedos among the X-types, are spread throughout the main belt regions,
and increases beyond 3 AU, while the proportion of medium-albedo M-types decreases.
P-type asteroids are considered to have similar origin or similar evolutional process, to C- or D-types. 

\acknowledgments

This study is based on observations with {\it AKARI}, a JAXA project with
the participation of ESA. We would like to thank all members of the
{\it AKARI} project for their dedicated efforts to generate this data base. 
This study was supported in part by ISAS/JAXA as a collaborative
program with the Space Plasma Experiment. This work is supported in part by 
a Grant-in-Aid for Scientific Research on Priority Areas No. 19047003 to DK, 
and Grants-in-Aid for Young Scientists (B) No. 21740153 and Scientific Research 
on Innovative Areas No. 21111005 to TO. 
Finally, we are very grateful to the anonymous reviewer, for her/his constructive comments.

\appendix

\section{Mean albedo for each taxonomic class}
Tables \ref{table:summary for Tholen types}, \ref{table:summary for Bus types}, 
and \ref{table:summary for Carvano types}
summarize the mean albedo of each taxonomic class, classified according to 
Tholen taxonomy ~\citep{Tholen1984, Lazzaro2004}, 
Bus taxonomy ~\citep{Bus1999, Lazzaro2004}, and Carvano taxonomy ~\citep{Carvano2010}, respectively.

\clearpage

\begin{deluxetable}{lrrrrrrr}
\tablecaption{Numbers of asteroids detected by {\it AKARI}, classified by dynamical group and 
by main belt region. \label{number of asteroids}}
\tablewidth{0pt}
\tablehead{Dynamical group       &       C &      S &       X &      D &        V & Unclassified & Total}
\startdata
Near-Earth Asteroids  &       2 &     16 &      2 &       -- &       1 &     37 & 58\\
Main Belt Asteroids    &    1150 &    722 &    490 &      130 &        4 &   2226 & 4722\\
Cybeles       &      22 &      2 &     26 &       26 &      -- &     29 & 105\\
Hildas        &       1 &      1 &     19 &       30 &      -- &     35 &  86\\
Jovian Trojans (L4, L5)       &       6 &     -- &      7 &       53 &      -- &     43 & 109\\
Others\tablenotemark{a}        &     2 &       17 &       6 &      2 &      -- &     13 & 40\\\hline
Total        &    1183 &    758 &    550 &      241 &       5 &   2383 & 5120\\\hline
\\
\multicolumn{8}{l}{MBAs subtotal}\\\hline
Inner    &    160 &     242 &      64 &      16 &       4 &     372 & 858\\
Middle   &    401 &     267 &     168 &      28 &      -- &     659 & 1523\\
Outer    &    589 &     213 &     258 &      86 &      -- &    1195 & 2341\\\hline
Total    &   1150 &     722 &     490 &     130 &       4 &    2226 & 4722\\
\\
\multicolumn{8}{l}{MBAs ($d > 20$ km) subtotal}\\\hline
Inner    &     48 &      56 &      32 &       8 &       1 &      11 &  156\\
Middle   &    184 &     138 &     124 &      17 &      -- &      61 &  524\\
Outer    &    381 &     147 &     212 &      62 &      -- &     492 & 1294\\\hline
Total    &    613 &     341 &     368 &      87 &       1 &     564 & 1974\\
\enddata
 \tablenotetext{a}{Others include asteroids belonging to the Hungaria and Thule families. }
\end{deluxetable}

\clearpage

\begin{deluxetable}{crrrlrl}
\tablecaption{Summary of the numbers ($N_{\rm ID}$) and mean albedos ($\overline{p_{\rm v}}$) in the five taxonomic types of total MBAs and MBAs larger than 20 km detected by {\it AKARI}. Same as table 8 in ~\citet{Usui2011}, but showing data for MBAs only. The number of 
asteroids with identified taxonomies is greater than that presented in ~\citet{Usui2011}, 
thanks to \citet{Carvano2010}. \label{table:summary for 5 types}}
\tablewidth{0pt}
\tablehead{
\multicolumn{1}{c}{\footnotesize type}     & %
\multicolumn{1}{c}{$N_{\rm ID}$} &%
\multicolumn{1}{c}{$\overline{p_{\rm v}}$}&%
\multicolumn{1}{c}{$N_{\rm ID} (d > 20 {\rm km})$} &%
\multicolumn{1}{c}{$\overline{p_{\rm v}} (d > 20 {\rm km})$}
}
\startdata
C     & 1150 & $0.071 \pm 0.040$ & 613 & $0.066 \pm 0.031$\\
S     &  722 & $0.208 \pm 0.079$ & 341 & $0.192 \pm 0.060$\\
X     &  490 & $0.098 \pm 0.081$ & 368 & $0.094 \pm 0.073$\\
D     &  130 & $0.086 \pm 0.053$ &  87 & $0.077 \pm 0.041$\\
V     &    4 & $0.297 \pm 0.131$ &   1 & $0.342$ \\\hline
total & 2496 &                   &1410 \\
\enddata
\end{deluxetable}

\clearpage

\begin{deluxetable}{llll}
\tablecaption{Mean albedo values ($\overline{p_{\rm v}}$) of the subclasses of X-type asteroids, separated by main belt region: 
inner, middle, and outer. Numbers of AcuA asteroids in each category are given 
in parentheses. \label{mean albedo for X-type}}
\tablewidth{0pt}
\tablehead{  & \multicolumn{1}{c}{E} & \multicolumn{1}{c}{M} & \multicolumn{1}{c}{P}
}
\startdata
Inner  & $0.454 \pm 0.119$ & $0.169 \pm 0.044$ & $0.063 \pm 0.017$\\
       & \multicolumn{1}{l}{(6)} & \multicolumn{1}{l}{(22)} & \multicolumn{1}{l}{(36)}\\
Middle & $0.397 \pm 0.076$ & $0.166 \pm 0.041$ & $0.057 \pm 0.018$\\
       & \multicolumn{1}{l}{(6)} & \multicolumn{1}{l}{(56)} & \multicolumn{1}{l}{(106)}\\
Outer  & $0.376 \pm 0.016$ & $0.166 \pm 0.050$ & $0.052 \pm 0.017$\\
       & \multicolumn{1}{l}{(2)} & \multicolumn{1}{l}{(67)} & \multicolumn{1}{l}{(189)}\\
\\
\multicolumn{4}{l}{$d > 20$ km}\\
\hline
Inner  & $0.479          $ & $0.161 \pm 0.033$ & $0.061 \pm 0.016$\\
       & \multicolumn{1}{l}{(1)} & \multicolumn{1}{l}{(10)} & \multicolumn{1}{l}{(21)}\\
Middle & $0.387 \pm 0.081$ & $0.165 \pm 0.039$ & $0.054 \pm 0.017$\\
       & \multicolumn{1}{l}{(5)} & \multicolumn{1}{l}{(42)} & \multicolumn{1}{l}{(77)}\\
Outer  & $0.365          $ & $0.169 \pm 0.050$ & $0.053 \pm 0.017$\\
       & \multicolumn{1}{l}{(1)} & \multicolumn{1}{l}{(58)} & \multicolumn{1}{l}{(153)}\\
\enddata
\end{deluxetable}

\clearpage

\begin{deluxetable}{rlllllll}
\tablecaption{
Large ($d > 20$ km) {E-type} asteroids.
\label{E-class asteroids}}
\tablewidth{0pt}
\tablehead{\multicolumn{2}{l}{Asteroid} & $d$ [km] & $p_{\rm v}$ & region & T\tablenotemark{a} & B\tablenotemark{b} & L\tablenotemark{c}}
\startdata
44 & Nysa    & $75.66 \pm 0.74$ & $0.479 \pm 0.013$ & inner  & E  & Xc & --  \\
55 & Pandora & $63.30 \pm 0.97$ & $0.337 \pm 0.013$ & middle & M  & X  & --  \\
64 & Angelina& $54.29 \pm 0.48$ & $0.515 \pm 0.012$ & middle & E  & Xe & --  \\
71 & Niobe   & $80.86 \pm 0.80$ & $0.326 \pm 0.008$ & middle & S  & Xe & --  \\
214& Aschera & $26.07 \pm 0.34$ & $0.419 \pm 0.013$ & middle & E  & Xc & B \\
504& Cora    & $30.39 \pm 0.35$ & $0.336 \pm 0.010$ & middle & -- & X  & X \\
665& Sabine  & $53.01 \pm 0.77$ & $0.365 \pm 0.012$ & outer  & -- & -- & X \\
\enddata
\tablenotetext{a}{~Tholen class~\citep{Tholen1984}.}
\tablenotetext{b}{~Bus class~\citep{Bus1999}.}
\tablenotetext{c}{~Lazzaro class~\citep{Lazzaro2004}.}
\end{deluxetable}

\clearpage

\clearpage

\clearpage

\clearpage

\clearpage

\clearpage

\clearpage

\clearpage

\clearpage

\clearpage

\clearpage

\clearpage

\begin{deluxetable}{ccrrr}
\tablecaption{
Summary of the numbers, mean albedos, and albedo variations of 784 asteroids detected by {\it AKARI}, 
classified according to Tholen taxonomy. 
The classification is based on the references in ~\citet{Usui2011}, appendix 4.
Note: Asteroids belonging to Q (S-type) and W (X-type) classes are not detected by {\it AKARI}.
\label{table:summary for Tholen types}
}
\tablewidth{0pt}
\tablehead{
type\tablenotemark{a}& class\tablenotemark{b}& %
\multicolumn{1}{c}{$N_{\rm ID}$}\tablenotemark{c}&%
\multicolumn{1}{c}{$\overline{p_{\rm v}}$}\tablenotemark{d}&%
\multicolumn{1}{c}{$\sigma\left(\overline{p_{\rm v}}\right)$\tablenotemark{e}}
}
\startdata
C & B           &      13  & 0.113  & 0.069\\
  & C           &     178  & 0.061  & 0.028\\
  & F           &      35  & 0.058  & 0.023\\
  & G           &      10  & 0.073  & 0.018\\\hline
S & A           &       5  & 0.282  & 0.101\\
  & R           &       1  & 0.277  & \multicolumn{1}{c}{---}\\
  & S           &     339  & 0.213  & 0.071\\
  & O           &       1  & 0.256  & \multicolumn{1}{c}{---}\\
  & K           &      23  & 0.143  & 0.039\\\hline
X & E           &       7  & 0.559  & 0.140\\
  & M           &      38  & 0.175  & 0.052\\
  & P           &      39  & 0.049  & 0.018\\\hline
D & D           &      84  & 0.061  & 0.030\\
  & T           &       7  & 0.075  & 0.023\\\hline
V & V           &       3  & 0.293  & 0.150\\
  & J           &       1  & 0.436  & \multicolumn{1}{c}{---}\\
\enddata
 \tablenotetext{a}{Taxonomic type of asteroids. }
 \tablenotetext{b}{Taxonomic class of asteroids determined by \citet{Tholen1984,Lazzaro2004}. }
 \tablenotetext{c}{Number of detections by {\it AKARI}. }
 \tablenotetext{d}{Mean geometric albedo in optical. }
 \tablenotetext{e}{Albedo variations. }
\end{deluxetable}

\clearpage

\begin{deluxetable}{ccrrr}
\tablecaption{
Summary of the numbers, mean albedos, and albedo variations of 894 asteroids detected by {\it AKARI}, 
classified according to Bus taxonomy.
Classification is based on the references in ~\citet{Usui2011}, appendix 4.
\label{table:summary for Bus types}
}
\tablewidth{0pt}
\tablehead{
type\tablenotemark{a} & class\tablenotemark{b} & %
\multicolumn{1}{c}{$N_{\rm ID}$}\tablenotemark{c}&%
\multicolumn{1}{c}{$\overline{p_{\rm v}}$}\tablenotemark{d}&%
\multicolumn{1}{c}{$\sigma\left(\overline{p_{\rm v}}\right)$\tablenotemark{e}}
}
\startdata
C & B            &      46  & 0.092  & 0.056\\
  & C            &     121  & 0.072  & 0.043\\
  & Cb           &      25  & 0.075  & 0.058\\
  & Cg           &       8  & 0.069  & 0.029\\
  & Cgh          &      11  & 0.097  & 0.041\\
  & Ch           &     128  & 0.063  & 0.023\\\hline
S & A            &       9  & 0.298  & 0.143\\
  & K            &      28  & 0.154  & 0.065\\
  & L            &      22  & 0.139  & 0.038\\
  & Ld           &       7  & 0.137  & 0.060\\
  & O            &       2  & 0.334  & 0.110\\
  & Q            &       2  & 0.353  & 0.001\\
  & R            &       1  & 0.277  & \multicolumn{1}{c}{---}\\
  & S            &     185  & 0.233  & 0.069\\
  & Sa           &      14  & 0.224  & 0.076\\
  & Sk           &      14  & 0.239  & 0.059\\
  & Sl           &      28  & 0.218  & 0.051\\
  & Sq           &      16  & 0.327  & 0.201\\
  & Sr           &       1  & 0.360  & \multicolumn{1}{c}{---}\\\hline
X & X            &     101  & 0.119  & 0.079\\
  & Xc           &      49  & 0.100  & 0.092\\
  & Xe           &      21  & 0.286  & 0.201\\
  & Xk           &      35  & 0.111  & 0.132\\\hline
D & D            &       7  & 0.086  & 0.056\\
  & T            &      11  & 0.054  & 0.009\\\hline
V & V            &       2  & 0.318  & 0.035\\
\enddata
 \tablenotetext{a}{Taxonomic type of asteroids. }
 \tablenotetext{b}{Taxonomic class of asteroids determined by \citet{Bus1999, Lazzaro2004}. }
 \tablenotetext{c}{Number of detections by {\it AKARI}. }
 \tablenotetext{d}{Mean geometric albedo in optical. }
 \tablenotetext{e}{Albedo variations. }
\end{deluxetable}

\clearpage

\begin{deluxetable}{ccrrr}
\tablecaption{
Summary of the numbers, mean albedos, and albedo variations of 1347 asteroids detected by {\it AKARI}, 
classified according to the Carvano taxonomy. 
Note: Asteroids belonging to CD, CO, CS, AV, LA, LQ, OV, QV, SO, SV, XS, DS
classes are not detected by {\it AKARI}.
\label{table:summary for Carvano types}
}
\tablewidth{0pt}
\tablehead{
type\tablenotemark{a} & class\tablenotemark{b} & %
\multicolumn{1}{c}{$N_{\rm ID}$}\tablenotemark{c}&%
\multicolumn{1}{c}{$\overline{p_{\rm v}}$}\tablenotemark{d}&%
\multicolumn{1}{c}{$\sigma\left(\overline{p_{\rm v}}\right)$\tablenotemark{e}}
}
\startdata
C & C  & 701 &0.069 &0.040\\
  & CL &   1 &0.149 &\multicolumn{1}{c}{---}\\
  & CQ &   1 &0.597 &\multicolumn{1}{c}{---}\\
  & CX &  66 &0.061 &0.034\\\hline
S & A  &   2 &0.268 &0.045\\
  & AQ &   1 &0.194 &\multicolumn{1}{c}{---}\\
  & L  &  80 &0.173 &0.091\\
  & LS &  32 &0.194 &0.059\\
  & O  &   3 &0.135 &0.115\\
  & Q  &   3 &0.269 &0.127\\
  & QO &   1 &0.054 &\multicolumn{1}{c}{---}\\
  & S  & 120 &0.218 &0.072\\
  & SA &   1 &0.232 &\multicolumn{1}{c}{---}\\
  & SQ &   2 &0.217 &0.027\\\hline
X & X  & 199 &0.084 &0.080\\
  & XD &  10 &0.080 &0.050\\
  & XL &   9 &0.123 &0.066\\\hline
D & D  & 111 &0.064 &0.026\\
  & DL &   3 &0.171 &0.065\\\hline
V & V  &   1 &0.2840 &\multicolumn{1}{c}{---}\\
\enddata
 \tablenotetext{a}{Taxonomic type of asteroids. }
 \tablenotetext{b}{Taxonomic class of asteroids determined by \citet{Carvano2010}. }
 \tablenotetext{c}{Number of detections by {\it AKARI}. }
 \tablenotetext{d}{Mean geometric albedo in optical. }
 \tablenotetext{e}{Albedo variations. }
\end{deluxetable}


\clearpage

\begin{thebibliography}{}

\bibitem[Allen(1970)]{Allen1970}
Allen, D. A.\ 1970, \nat, 227, 158

\bibitem[Allen(1971)]{Allen1971}
Allen, D. A.\ 1971, 
in Physical Studies of Minor Planets, 
ed.\  T. Gehrels (Washington: National Aeronautics and Space Administration SP-267), 41

\bibitem[Bell et al.(1989)]{Bell1989}
Bell, J. F., et al.\ 1989, 
in Asteroids II, ed.\  R. P. Binzel et al. (Tucson: University of Arizona Press), 921

\bibitem[Binzel et al.(2010)]{Binzel2010}
Binzel, R. P., et al.\ 2010, \nat, 463, 331

\bibitem[Binzel et al.(2004)]{Binzel2004}
Binzel, R. P., Rivkin, A. S., Stuart, J. S., Harris, A. W., Bus, S. J., \& Burbine, T. H.\ 2004, \icarus, 170, 259

\bibitem[Bottke et al.(2005)]{Bottke2005}
Bottke, W. F., Jr., Durda, D. D.,  Nesvorn\'{y}, D., Jedicke, R., Morbidelli, A., Vokrouhlick\'{y}, D., \& Levison, H. F.\ 2005, \icarus, 179, 63

\bibitem[Bottke et al.(2002a)]{Bottke2002a}
Bottke, W. F., Jr., Cellino, A., Paolicchi, P., \& Binzel, R. P.\ 2002, 
in Asteroids III, ed.\  W. F. Bottke et al. (Tucson: University of Arizona Press), 3

\bibitem[Bottke et al.(2002b)]{Bottke2002b}
Bottke, W. F., Jr., Vokrouhlicky\'{y}, D., Rubincam, D. P., \& Broz, M.\ 2002, 
in Asteroids III, ed.\  W. F. Bottke et al. (Tucson: University of Arizona Press), 395

\bibitem[Bowell et al.(1994)]{Bowell1994}	
Bowell, E., Muinonen, K., \& Wasserman, L. H.\ 1994, 
in Asteroids, Comets, Meteors 1993, 
ed.\  A. Milani et al. (Dordrecht: Kluwer Academic Publishers), 477

\bibitem[Brownlee et al.(2003)]{Brownlee2003}
Brownlee, D. E., et al.\ 2003, 
Journal of Geophysical Research, 108, 8111

\bibitem[Burbine et al.(2008)]{Burbine2008}
Burbine, T. H., Rivkin, A. S., Noble, S. K., Moth\'{e}-Diniz, T., Bottke, W. F., McCoy, T. J.,
Dyar, M. D., \& Thomas, C.A.\ 2008, in Reviews in Mineralogy \& Geochemistry 68, 273

\bibitem[Bus(1999)]{Bus1999}
Bus, S. J.\ 1999, PhD thesis, Massachusetts Institute of Technology

\bibitem[Bus \& Binzel(2002)]{Bus2002}	
Bus, S. J., \& Binzel, R. P.\ 2002, \icarus, 158, 146

\bibitem[Carvano et al.(2010)]{Carvano2010}
Carvano, J. M., Hasselmann, P. H., Lazzaro, D., \& Moth\'{e}-Diniz, T.\ 2010, \aap, 510, A43

\bibitem[Carvano et al.(2003)]{Carvano2003}
Carvano, J. M., Moth\'{e}-Diniz, T., \& Lazzaro, D.\ 2003, \icarus, 161, 356

\bibitem[Chapman et al.(1975)]{Chapman1975}
Chapman, C. R., Morrison, D., \& Zellner, B.\ 1975, \icarus, 25, 104

\bibitem[Cheng et al.(1997)]{Cheng1997}
Cheng, A. F., et al.\ 1997, 
Journal of Geophysical Research, 102, 23695

\bibitem[Clark et al.(2004)]{Clark2004}
Clark, B. E., et al. 2004, \aj, 128, 3070

\bibitem[Clark et al.(2009)]{Clark2009}
Clark, B. E., et al. 2009, \icarus, 202, 119

\bibitem[Davis et al.(1979)]{Davis1979}	
Davis, D. R., Chapman, C. R., Greenberg, R., Weidenschilling, S. J., \&  Harris, A. W.\ 1979, 
in Asteroids (Tucson: University of Arizona Press), 528

\bibitem[Davis et al.(2002)]{Davis2002}
Davis, D. R., Durda, D. D., Marzari, F., Campo Bagatin, A., \& Gil-Hutton, R.\ 2002, 
in Asteroids III, ed.\  W. F. Bottke et al. (Tucson: University of Arizona Press), 545

\bibitem[Drummond et al.(1985)]{Drummond1985}
Drummond, J. D., Cocke, W. J., Hege, E. K., \& Strittmatter, P. A.\ 1985, \icarus, 61, 132

\bibitem[Drummond et al.(2009)]{Drummond2009}
Drummond, J., Christou, J., \& Nelson, J.\ 2009, \icarus, 202, 147

\bibitem[Durech et al.(2011)]{Durech2011}
\u{D}urech, J., et al.\ 2011, \icarus, 214, 652

\bibitem[Fornasier et al.(2011)]{Fornasier2011}
Fornasier, S., Clark, B. E., \& Dotto, E.\ 2011, \icarus, 214, 131

\bibitem[Fowler \& Chillemi(1992)]{Fowler1992}
Fowler, J. W., \& Chillemi, J. R.\ 1992, 
Phillips Lab. Tech. Rep., 2049, 17

\bibitem[Froeschle \& Greenberg(1989)]{Froeschle1989}	
Froeschle, Cl., \& Greenberg, R.\ 1989, 
in Asteroids II, ed.\  R. P. Binzel et al. (Tucson: University of Arizona Press), 827

\bibitem[Fujiwara(1982)]{Fujiwara1982}	
Fujiwara, A.\ 1982, \icarus, 52, 434

\bibitem[Fujiwara et al.(2006)]{Fujiwara2006}
Fujiwara, A., et al.\ 2006, Science, 312, 1330

\bibitem[Gaffey et al.(2002)]{Gaffey2002}
Gaffey, M. J., Cloutis, E. A., Kelley, M. S.,\& Reed, K. L.\ 2002, 
in Asteroids III, ed.\  W. F. Bottke et al. (Tucson: University of Arizona Press), 138

\bibitem[Glassmeier et al.(2007)]{Glassmeier2007}
Glassmeier, K.-H., Boehnhardt, H., Koschny, D., K\"{u}hrt, E., \& Richter, I.\ 2007, 
Space Science Reviews, 128, 1

\bibitem[Gradie et al.(1979)]{Gradie1979}
Gradie, J. C., Chapman, C. R., \& Williams, J. G.\ 1979, 
in Asteroids (Tucson: University of Arizona Press), 359

\bibitem[Greenwood et al.(2010)]{Greenwood2010}
Greenwood, R. C., Franchi, I. A., Kearsley, A. T., \& Alard, O.\ 2010, 
Geochimica et Cosmochimica Acta, 74, 1684

\bibitem[Hirata \& Ishiguro(2011)]{Hirata2011}
Hirata, N., \& Ishiguro, M.\ 2011, 42nd Lunar and Planetary Science Conference, 1821

\bibitem[Hiroi \& Pieters(1992)]{Hiroi1992}
Hiroi, T., \& Pieters, C. M.\ 1992, in Proc. Lunar and Planetary Science, 22, 313
	
\bibitem[Hiroi \& Pieters(1994)]{Hiroi1994a}
Hiroi, T., \& Pieters, C. M.\ 1994, Journal of Geophysical Research, 99, 10867

\bibitem[Hiroi et al.(1994)]{Hiroi1994}
Hiroi, T.,  Pieters, C. M.,  Zolensky, M. E., \& Lipschutz, M. E.\ 1994, 
in Proceedings NIPR Symposium No. 7, ed.\ K. Yanai, et al. (National Institute of Polar Research), 230

\bibitem[Hiroi et al.(2001)]{Hiroi2001}
Hiroi, T., Zolensky, M. E., \& Pieters, C. M.\ 2001, Science, 293, 2234

\bibitem[Ishihara et al.(2010)]{Ishihara2010}
Ishihara, D., et al.\ 2010, \aap, 514, A1

\bibitem[Johnson et al.(1992)]{Johnson1992}
Johnson, T. V., Yeates, C. M., \& Young, R.\ 1992, 
Space Science Reviews, 60, 3

\bibitem[Kasuga et al.(2012)]{Kasuga2012}
Kasuga, T., et al.\ 2012, \aj, 143, 141

\bibitem[Kirkwood(1867)]{Kirkwood1867}
Kirkwood, D.\ 1867, Meteoric astronomy: a treatise on shooting-stars, fireballs, and aerolites., 
Philadelphia, J. B. Lippincott \& co.

\bibitem[Lagerkvist et al.(2005)]{Lagerkvist2005}
Lagerkvist, C.-I., Moroz, L., Nathues, A., Erikson, A., Lahulla, F., Karlsson, O., \& Dahlgren, M.\ 2005, \aap, 432, 349

\bibitem[Lazzarin et al.(2006)]{Lazzarin2006}
Lazzarin, M., Marchi, S., Moroz, L. V., Brunetto, R., Magrin, S., Paolicchi, P., \& Strazzulla, G.\ 2006, \apj, 647, L179

\bibitem[Lazzaro et al.(2004)]{Lazzaro2004}
Lazzaro, D., Angeli, C. A., Carvano, J. M., Moth\'{e}-Diniz, T., Duffard, R., \& Florczak, M.\ 2004, \icarus, 172, 179

\bibitem[Lebofsky et al.(1986)]{Lebofsky1986}
Lebofsky, L. A., et al.\ 1986, \icarus, 68, 239

\bibitem[Levison et al.(2009)]{Levison2009}
Levison, H. F., Bottke, W. F., Gounelle, M., Morbidelli, A., Nesvorn\'{y}, D., \& Tsiganis, K.\ 2009, \nat, 460, 364

\bibitem[Li et al.(2010)]{Li2010}
Li, J.-Y., et al.\ 2010, \icarus, 208, 238

\bibitem[Mainzer et al.(2011)]{Mainzer2011}
Mainzer, A., et al.\ 2011, \apj, 731, 53

\bibitem[Masiero et al.(2011)]{Masiero2011}
Masiero, J. R., et al.\ 2011, \apj, 741, 68

\bibitem[Matson(1971)]{Matson1971}
Matson, D. L.\ 1971, 
in IAU Colloq. 12, Physical Studies of Minor Planets, 
ed.\  T. Gehrels (Washington: National Aeronautics and Space Administration SP-267), 45

\bibitem[Micha{\l}owski et al.(2006)]{Michalowski2006}
Micha{\l}owski, T., et al.\ 2006, \aap, 459, 663

\bibitem[Morbidelli et al.(2002)]{Morbidelli2002}
Morbidelli, A., Jedicke, R., Bottke, W. F., Michel, P., \& Tedesco, E. F.\ 2002, \icarus, 158, 329
	
\bibitem[Moth\'{e}-Diniz(2009)]{Mothe-Diniz2009}
Moth\'{e}-Diniz, T.\ 2009, 
Icy Bodies of the Solar System, Proceedings IAU Symposium No. 263, 
ed. J.A. Fernaandez, et al., 231

\bibitem[Moth\'{e}-Diniz et al.(2003)]{Mothe-Diniz2003}
Moth\'{e}-Diniz, T., Carvano, J. M., \& Lazzaro, D.\ 2003, \icarus, 162, 10

\bibitem[Murakami et al.(2007)]{Murakami2007}
Murakami, H., et al.\ 2007, \pasj, 59, S369

\bibitem[Neese(2010)]{Neese2010}
Neese, C., Ed., Asteroid Taxonomy V6.0. EAR-A-5-DDR-TAXONOMY-V6.0. NASA Planetary Data System, 2010.

\bibitem[Nesvorn\'{y} et al.(2005)]{Nesvorny2005}
Nesvorn\'{y}, D., Jedicke, R., Whiteley, R. J., \& Ivezi\'{c}, \u{Z}.\ 2005, \icarus, 173, 132

\bibitem[Neugebauer et al.(1984)]{Neugebauer1984}
Neugebauer, G., et al.\ 1984, \apj, 278, L1

\bibitem[Onaka et al.(2007)]{Onaka2007}
Onaka, T., et al.\ 2007, \pasj, 59, S401

\bibitem[Ostro et al.(2000)]{Ostro2000}
Ostro, S. J., et al.\ 2000, Science, 288, 836

\bibitem[Pravec et al.(2012)]{Pravec2012}
Pravec, P., et al.\ 2012, \icarus, 221, 365

\bibitem[Rayman(2003)]{Rayman2003}
Rayman, M. D.\ 2003, 
Space Technology 23, 185

\bibitem[Russell et al.(2004)]{Russell2004}
Russell, C. T., et al.\ 2004, 
Planetary and Space Science, 52, 465

\bibitem[Sasaki et al.(2001)]{Sasaki2001}
Sasaki, S., Nakamura, K., Hamabe, Y., Kurahashi, E., \& Hiroi, T.\ 2001, \nat, 410, 555

\bibitem[Scholl et al.(1989)]{Scholl1989}	
Scholl, H., et al.\ 1989, 
in Asteroids II, ed.\  R. P. Binzel et al. (Tucson: University of Arizona Press), 845

\bibitem[Stern \& Spencer(2003)]{Stern2003}
Stern, A., \& Spencer, J.\ 2003, 
Earth, Moon, and Planets, 92, 477

\bibitem[Tedesco et al.(2002)]{Tedesco2002}
Tedesco, E. F., Noah P. V., Noah, M., \& Price, S. D.\ 2002, \aj, 123, 1056

\bibitem[Tholen(1984)]{Tholen1984}
Tholen, D. J.\ 1984, PhD thesis, Arizona University

\bibitem[Tholen \& Barucci(1989)]{Tholen1989}	
Tholen, D. J., \& Barucci, M. A.\ 1989, 
in Asteroids II, ed.\  R. P. Binzel et al. (Tucson: University of Arizona Press), 298
	
\bibitem[Thomas et al.(2011)]{Thomas2011}
Thomas, C. A., Rivkin, A. S., Trilling, D. E., Marie-Therese E., \& Grier, J. A.\ 2011, \icarus, 212, 158
	
\bibitem[Trieloff et al.(2003)]{Trieloff2003}
Trieloff, M., et al. 2003, \nat, 422, 502

\bibitem[Usui et al.(2011)]{Usui2011}
Usui, F., et al.\ 2011, \pasj, 63, 1117

\bibitem[Warner et al.(2009)]{Warner2009}
Warner, B. D., et al.\ 2009, \icarus, 204, 172

\bibitem[Weisberg et al.(2006)]{Weisberg2006}
Weisberg, M. K., McCoy, T. J., \& Krot, A. N.\ 2006, 
in Meteorites and the Early Solar System II, ed.\ D.S. Lauretta \& H.Y. McSween 
(Tucson: University of Arizona Press), 19
	
\bibitem[Wright et al.(2010)]{Wright2010}
Wright, E. L., et al.\ 2010, \aj, 140, 1868

\bibitem[Yoshikawa(1989)]{Yoshikawa1989}
Yoshikawa, M.\ 1989, \aap, 213, 436

\bibitem[Zellner \& Gradie(1976)]{Zellner1976}
Zellner, B., \& Gradie, J.\ 1976, \aj, 81, 262

\bibitem[Zellner(1979)]{Zellner1979}
Zellner, B.\ 1979, 
in Asteroids (Tucson: University of Arizona Press), 783

\end{thebibliography}
\end{document}